\documentclass{aa}
\usepackage{natbib}
\bibpunct{(}{)}{;}{a}{}{,}

\usepackage{graphicx}
\usepackage{color}
\usepackage{amsmath}
\usepackage{txfonts}
\usepackage{hyperref}   
\hypersetup{colorlinks=true,linkcolor=[rgb]{1.,0.2,0.2},citecolor=[rgb]{0.1,0.4,1.},filecolor=[rgb]{0.7,0.2,0.2},urlcolor=[rgb]{0.7,0.2,0.2}}

\usepackage{arydshln}
\usepackage{multirow}

\definecolor{darkspringgreen}{rgb}{0.09, 0.45, 0.27}
\definecolor{regalia}{rgb}{0.32, 0.18, 0.5}

\newcommand{\orcid}[1]{\href{https://orcid.org/#1}{\textcolor[HTML]{A6CE39}{\aiOrcid}}}

\begin{document} 

   \title{Mixture Models for Photometric Redshifts}

  \author{
          Z. Ansari
          \inst{1}\fnmsep\thanks{ORCID 0000-0002-4775-9685}
          \and A. Agnello
          \inst{1}\fnmsep\thanks{ORCID 0000-0001-9775-0331}
          \and C. Gall 
          \inst{1}\fnmsep\thanks{ORCID 0000-0002-8526-3963}
          }
   \authorrunning{Ansari, Agnello, Gall}

   \institute{$^1$DARK, Niels Bohr Institute, University of Copenhagen,
              Jagtvej 128, 2200 Copenhagen, Denmark\\
              }

\abstract
{Determining photometric redshifts (photo-$z$s) of extragalactic sources
to high accuracy is paramount to measure distances in wide-field cosmological experiments. With only photometric information at hand, photo-$z$s are prone to systematic uncertainties in the intervening extinction and the unknown underlying spectral-energy distribution of different astrophysical sources, leading to degeneracies in modern machine learning algorithm that impact the level of accuracy for photo-$z$ estimates.}
{Here, we aim to resolve these model degeneracies and obtain a clear separation between intrinsic physical properties of astrophysical sources and extrinsic systematics. Furthermore, we aim at meaningful estimates of the full photo-$z$ probability distributions, and their uncertainties.}
{We perform a probabilistic photo-$z$ determination using Mixture Density Networks (MDN). The training data-set is composed of optical ($griz$ photometric bands) point-spread-function and model magnitudes and extinction measurements from the SDSS-DR15, and {\it WISE} mid-infrared (3.4$\mu$m and 4.6$\mu$m) model magnitudes.
We use Infinite Gaussian Mixture models to classify the objects in our data-set as stars, galaxies or quasars, and to determine the number of MDN components to achieve optimal performance.}
{The fraction of objects that are correctly split into the main classes of stars, galaxies and quasars is 94\%. Furthermore, our method improves the bias of photometric redshift estimation (i.e. the mean $\Delta z =(z_{p} - z_{s})/(1+ z_{s})$) by one order of magnitude compared to the SDSS photo-$z$, and decreases the fraction of $3\sigma$ outliers (i.e. $3 \times rms(\Delta z) < \Delta z$). The relative, root-mean-square systematic uncertainty in our resulting photo-$z$s is down to 1.7\% for benchmark samples of low-redshift galaxies ($z_{s}<0.5$).}
{We have demonstrated the feasibility of machine-learning based methods that produce full probability distributions for photo-$z$ estimates with a performance that is competitive with state-of-the art techniques. Our method can be applied to wide-field surveys where extinction can vary significantly across the sky and with sparse spectroscopic calibration samples.}


   \maketitle

\section{Introduction}
The redshift of an astrophysical object is routinely determined from absorption or emission lines in its spectrum. In the absence of spectroscopic information, its \textit{photometric redshift} (hereafter photo-$z$) can be estimated from the apparent luminosity measured in different photometric bands \citep[see e.g.][for a general review]{Salvato2019}. Accurate photo-$z$s are needed by wide-field surveys that seek to probe cosmology through the spatial correlations of the matter density field, and are in fact a core limiting factor in the accuracy of these measurements \citep[e.g.,][]{2006ApJ...652..857K}.

While large areas of the sky are covered by optical and near-IR imaging surveys, only a minority of objects have observed spectra -- and hence secure redshifts from emission or absorption lines. The major problem is the rather narrow wavelength range covered by most photometric bands that introduces uncertainties and degeneracies in the redshift estimation. Some photo-$z$ {calibration fields} exist, with extensive spectroscopic campaigns (albeit with some non-negligible pre-selection) and moderately deep photometry in the optical and near infrared (NIR), covering a few square degrees of sky in total. Notably, the PRIMUS \citep[][]{2011ApJ...741....8C, 2013ApJ...767..118C} and zCOSMOS \citep{2008Msngr.134...35L} have been used by the Kilo-Degree Survey Collaboration \citep[KiDS;][]{2013ExA....35...25D} and Dark Energy Survey Collaboration \citep[DES;][]{2018ApJS..239...18A}, for the measurement of the matter content ($\Omega_{m}$) and present-day root-mean-square (rms) matter density fluctuations ($\sigma_{8}$). \citet{Hildebrandt17} have identified the different calibrations of photo-$z$s, across PRIMUS and zCOSMOS, to explain the difference in inferred cosmological parameters between DES and KiDS, claiming that the uncertainties in photo-$z$s are one outstanding challenge towards percent-level cosmology from weak lensing.

When only photometric information is available, a \textit{three-fold degeneracy} between an object type, its redshift, and foreground extinction hinders the unambiguous determination of the redshift. \citet{Galametz17} have quantified this effect explicitly in view of a possible synergy between the ESA-\textit{Euclid} mission
\citep{2012SPIE.8442E..0ZA}
and \textit{Rubin}-Legacy Survey of Space and Time
\citep[LSST;][]{2012SPIE.8442E..0ZA},
which should cover more than half of the extragalactic sky to $\gtrsim24$ mag depth in $YJH$-bands and $ugriz$-bands, respectively.

Here, we explore a probabilistic approach to compute photo-$z$s that account for the existence of an indefinite number of astrophysical object types and their cross-contamination due to broad-band imaging information. Specifically, we train a suite of Mixture-Density Networks \citep[MDNs,][]{astonpr373} to predict the probability distribution of the photo-$z$ of an object with measured magnitudes in multiple photometric bands as well as Galactic extinction. Following the standard nomenclature of Machine-Learning works, we will alternatively refer to the photometric properties (magnitudes and extinction) as \textit{features} in the rest of this paper. 
The MDN output is a sum of Gaussian functions in photo-$z$, whose parameters (i.e. the average, dispersion, amplitude) are non-linear combinations of the photometric inputs such as magnitude and extinction. Throughout the paper, we will term these output Gaussians as \textit{branches}.
In order to determine the number of branches that are needed to optimally parameterize the photo-$z$ probabilities, we must determine the range of MDN branches that will most accurately describe the data-set. Hence, we explore Infinite Gaussian Mixture Models (IGMM) on a
photometric sample of which about 2\% of the sources have spectroscopic redshifts (see sect.~\ref{SS:data}).
\begin{table*}
\caption{Recent automated approaches to estimate photo-$z$s.}
\label{table:MLs}      
\centering          
\begin{tabular}{p{1.3cm}|p{2.2cm}|p{1.8cm}|p{1.2cm}|p{2.2cm}|p{2.3cm}|p{3.9cm}} 
\hline\hline       
  Reference & Method\tablefootmark{a} & Photometric information & Objects & $z_{s}$ range\tablefootmark{b} & Depth [mag]\tablefootmark{c} & Survey 
  \\
\hline
    1 & kNN & $ugrizy$\tablefootmark{d} & Galaxies & $0 < z \leq 2$ & $i < 25.3$ & mock galaxy for LSST from DESC
    \\
    \hline
    2
    & ANN & $ugriz$\tablefootmark{e}, $E(B-V)$ & Galaxies & $z<0.4$ & $r_{Petro} \leq 17.8$ & SDSS-DR12  \\
    \hline
    3
    & METAPHOr,  & $ugri$ GAaP & Galaxies & $z_{s} \leq 1$ &  $r \leq 21$ & SDSS-DR9, 
    \\
    &ANN,  & & & & & KiDS ESO-DR3, \\
    &template fitting & & & & & GAMA-DR2, 2dFGRS
    \\
    \hline
    4
    & ANN & $ugriz$ \tablefootmark{e} & Galaxies & $z_{s} \leq 0.4$ &  $r_{Petro} \leq 17.8$ & SDSS/BOSS-DR12, GAMA-DR3
    \\
    \hline
    5 & kNN & $UV$, $ugrizy$, 
    $YJHK$  & Galaxies &  $0.3 < z_{s} < 3.0$ & $i < 25$ & mock galaxy catalogs for \textit{Euclid}, RST,\\
    & & & & & & and/or CASTOR
    \\
    \hline
    6 & kNN & $UV$, $ugriz$, $w1w2w3w4$ & QSOs  & --- & $14.7<r<22.6$ & SDSS-DR12, 2MASS, WISE 
    \\
    \hline
    7 & kNN & $grizy$ & Galaxies & $z_{s}>0.01$  & $18.5<i<25$ &SDSS/BOSS-DR14, DEEP2/3DR4, 
    \\
     & & & & & & VANDELS-DR2,\\
     & & & & & & COSMOS, C3R2,\\
     & & & & & & COSMOS2015
    \\
    \hline
    8 & tree based & $ugriz$, $BRI$ & Galaxies &  $0.02 \leq z_{s} \leq 0.3$ & $B_{AB}<$24.1 & SDSS/MGS-DR7,\\
     & & & & & & DEEP2-DR4
    \\
    \hline
    9 & tree based & $ugriz$ & Galaxies & $z_{s} \leq 0.55$ &$r_{Petro} < 17.77$ & SDSS-DR6, 2dF-SDSS LRG, 2SLAQ, DEEP2
    \\
    \hline
    10 & Gaussian process & $griz$, $RIZ$, $YJH$ & Galaxies & $0 \leq z_{s} \leq 2 $ & $RIZ< 25 $ & SDSS/BOSS
    \\
    \hline
    11 & ensemble of & $ugriz$ & Galaxies &$z_{s}<0.8$ & $i_{AB}\lesssim 22.5$ & SDSS/BOSS-DR10\\
     & ANNs, trees & & & & &\\
     & and kNN & & & & &\\
    \hline
    12 & ANN,& $grizy$\tablefootmark{f}, & Galaxies, & $z_{s}<1.5$ & $i \lesssim 23.1$ & PS1 $3\pi$ DR1, \\
    & Monte-Carlo, extrapolation& $E(B-V)$\tablefootmark{g} & QSOs, Stars & & & SDSS-DR14, DEEP2-DR4, VIPERS PDR-2, WiggleZ, zCOSMOS-DR3, VVDS\\
\hline
\end{tabular}
\tablebib{(1)~\citet{Schmidt2020}; (2)~\citet{Pasquet2019}; (3)~\cite{2019MNRAS.482.3116A}; (4)~\cite{Shuntov2020}; (5)~ \cite{2018AJ....155....1G}; (6)~\cite{2020MNRAS.493L..70C}; (7)~\cite{2020arXiv200301511N}; (8)~\cite{2013MNRAS.432.1483C}; (9)~\cite{2010ApJ...715..823G}; (10)~\cite{2016MNRAS.455.2387A}; (11)~\cite{2019ascl.soft10014S};
(12)~\cite{2020MNRAS.tmp.2033B}}
\tablefoot{
\tablefoottext{a}{
METAPHOr (Machine-learning Estimation Tool for Accurate PHOtometric Redshifts)
;}
\tablefoottext{b}{Spectroscopic redshift range.
;}
\tablefoottext{c}{Petrosian $r$-band magnitude, $r_{Petro}$; }
\tablefoottext{d}{Grey scale $48 \times 48$ pixel images; }
\tablefoottext{e}{Images in $ugriz$, $64\times64$ piexels in each band; }
\tablefoottext{f}{Magnitudes for PSF, Kron and seeing-matched apertures (FPSFMag, FKronMag and FApMag, respectively), as well as 3.00'', 4.63'' and 7.43'' fixed-radius apertures (FmeanMagR5, FmeanMagR6 and FmeanMagR7); 
}
\tablefoottext{g}{PS1 and Planck extinction maps.}
}
\end{table*}
\begin{table*}
\caption{Comparison of photo-$z$ estimates.}
\label{table:comparison}      
\centering          
\begin{tabular}{p{1.3cm}|p{8cm}|p{2cm}|p{2.5cm}|p{1.8cm}} 
  \hline\hline
 Reference & Method\tablefootmark{a}  & Bias\tablefootmark{b} & rms\tablefootmark{c} & Fraction of outlier in \% 
  \\
 \hline
  1 & (trainZ) &  $-$0.2086  & 0.1808 & 0    \\
    & ANNz2    &  0.00063    & 0.0270 & 4.4  \\
    & BPZ      &  $-$0.00175 & 0.0215 & 3.5  \\
    & Delight  &  $-$0.00185 & 0.0212 & 3.8  \\
    & EAZY     &  $-$0.00218 & 0.0225 & 3.4  \\
    & FlexZBoost & $-$0.00027 & 0.0154 & 2.0  \\
    & GPz      &  0.00000    & 0.0197 & 5.2  \\
    & Lephare  &  $-$0.00161 & 0.0236 & 5.8  \\
    & METAPhoR &  0.00000    & 0.0264 & 3.7  \\
    & CMNN     &  $-$0.00132 & 0.0184 & 3.5  \\
    & SkyNet   &  $-$0.00167 & 0.0219 & 3.6  \\
    & TPZ      &  0.00309    & 0.0161 & 3.3  \\
 \hline
  2  & Convolutional neural network(CNN) &  0.0001 & 0.0456\tablefootmark{d} & 0.31 \\
 \hline
  3 & METAPHOR &  $-$0.004 & 0.065 & 0.98 \\
    & ANNz2    &  $-$0.008 & 0.078 & 1.60 \\
    & BPZ      &  $-$0.020 & 0.048 & 1.13 \\
 \hline
  4  &  CNN + density field (mode) & 0.0038\tablefootmark{d} & & 0.83 \\
    & CNN + density field (median) & 0.0045\tablefootmark{d} & --- & 0.44  \\
    & CNN + density field (mean) &  0.0066\tablefootmark{d} & & 0.31 \\
 \hline
  5  & kNN & $-0.0001\pm 0.0$& $0.0165 \pm 0.0001 $  &  4.0 \\
 \hline
  6  & kNN &   0.001\tablefootmark{e} & 0.36 & 10.7\tablefootmark{f}\\
 \hline
 7 & DEmP\tablefootmark{g} &  -0.0291 &  0.1018 &  0.16 
 \\
  & DEmP\tablefootmark{h} & -0.0175& 0.07 & 0.17 
 \\
 \hline
 8 & Trees and Random Forest(Regression mode) &  -0.00008 & 0.0225 & 0  \\

  & Trees and Random Forest (Classification mode) &  0.00218 & 0.0246 & 0  \\ 
 \hline
  9 & ArborZ & -0.006\tablefootmark{e} & 0.985 & 1.9 \\
 \hline
  10 & GP-GL &  0.0946 & 0.1420 & 5.3  \\
   & GP-VL &  0.828  & 0.1251 & 5.5  \\  
   & GP-VC &  0.0294 & 0.0435 & 4.7    
   \\ 
 \hline
 11 &ensemble of ANNs, trees and KNN (nominal solution) &  0.0002 & 0.034 &0.105  \\
  & ensemble of ANNs, trees and KNN($<PDF>$) &0.00035 & 0.034 & 0.105  \\
  & ensemble of ANNs, trees and KNN(PDF) &  0.00035 & 0.052 & 0.1\\
  \hline
  12 & PS1-STRM  (All validation) base estimate & 0.0003 & 0.0342 & 2.88 \tablefootmark{i} \\
  & PS1-STRM  (All validation) Monte-Carlo sampled & 0.0010 & 0.0344 & 2.99 \\
  & PS1-STRM  (Non-extrapolated) base estimate
  & 0.0005  & 0.0322 & 1.89 \\
  & PS1-STRM  (Non-extrapolated) Monte-Carlo sampled & 0.0013 & 0.0323 & 2.00 \\
  \hline
\end{tabular}
\tablebib{(1)~\citet{Schmidt2020}; (2)~\citet{Pasquet2019}; (3)~\cite{2019MNRAS.482.3116A}; (4)~\cite{Shuntov2020}; (5)~\cite{2018AJ....155....1G}; (6)~\cite{2020MNRAS.493L..70C}; (7)~\cite{2020arXiv200301511N}; (8)~\cite{2013MNRAS.432.1483C}; (9)~\cite{2010ApJ...715..823G}; (10)~\cite{2016MNRAS.455.2387A};
(11)~\cite{2019ascl.soft10014S};
(12)~\cite{2020MNRAS.tmp.2033B}
}
\tablefoot{
{Values are provided where information was available.}
\tablefoottext{a}{Acronyms are defined in the respective literature; }
\tablefoottext{b}{Bias: defined as mean of $\Delta z = (z_{p} - z_{s})/(1+ z_{s})$; }
\tablefoottext{c}{rms($(z_{p} - z_{s})/(1+ z_{s})$); }
\tablefoottext{d}{$\sigma_{MAD} = 1.4826 \times MAD$, where MAD (Median Absolute Deviation) is the median of $|\Delta z - Median(\Delta z)|$; }
\tablefoottext{e}{Average of ${\delta z} =z_{p} - z_{s}$; }
\tablefoottext{f}{Fraction of outliers defined as number of objects with $|\Delta z| > rms(\Delta z) \pm 0.5$; }
\tablefoottext{g}{Exclusively using wide-band photometry from \textit{Wide} fields of HSC (https://hsc.mtk.nao.ac.jp/ssp/) as additional photometric input; }
\tablefoottext{h}{Exclusively using deep photometry from \textit{Deep} and \textit{UltraDeep} fields of HSC as additional photometric input; }
\tablefoottext{i}{Fraction of outliers defined as number of objects with $|\Delta z|> 0.15$ .}
}
\end{table*}
\subsection{Photometric Redshifts in the Literature}
\label{ss:photoz-in-lit}
There are two main methods commonly used to estimate photometric redshifts: (\textsc{i}) template fitting and (\textsc{ii}) machine learning algorithms. Template fitting methods specify the relation between synthetic magnitudes and redshift with a suite of spectral templates across a range of redshifts and object classes, through maximum likelihood \citep[e.g.][]{1999ApJ...513...34F} or Bayesian techniques \citep[e.g.][]{2000ApJ...536..571B, 2008ApJ...686.1503B, 2006A&A...457..841I}. Machine learning methods, using either images or a vector of magnitudes and colours, learn the relation between magnitude and redshift from a training data-set of objects with known spectroscopic redshifts. In principle, template fitting techniques do not require a large sample of objects with spectroscopic redshifts for training, and can be applied to different surveys and redshift coverages. However, these methods are computationally intensive and require explicit assumptions on e.g. dust extinction, which can lead to a degeneracy in colour-redshift space. Moreover, template fitting techniques are only as predictive as the family of available templates. In the case of large  samples of objects with spectroscopic redshifts, machine learning approaches such as artificial neural networks \citep[ANNs; e.g.][]{2019MNRAS.482.3116A, Shuntov2020}, k-nearest neighbours \citep[kNN; e.g.][]{2020MNRAS.493L..70C, 2018AJ....155....1G, 2020arXiv200301511N}, tree-based algorithms \citep[e.g.][]{2013MNRAS.432.1483C, 2010ApJ...715..823G} or Gaussian processes \citep[e.g.][]{2016MNRAS.455.2387A} have shown similar or better performances than the template fitting methods. However, machine learning algorithms are only reliable in the range of input values of their training data-set. Additionally, a lack of sufficient high-redshift spectroscopic samples affects the performance of machine learning implementations on photo-$z$ estimates. Another aspect is the production of photo-$z$ probability distributions given the photometric measurements: while template-based methods can easily produce a probability distribution by combining likelihoods from different object templates, most of the machine-learning methods in the literature are only trained to produce point estimates, i.e. just one photo-$z$ value for each object. 
For the sake of completeness, we summarise the state-of-the-art (and heterogeneous) efforts in the literature in Table~\ref{table:MLs}, and their performance metrics evaluation in Table~\ref{table:comparison}. We emphasize that most of the photo-$z$ estimation methods above have been trained and tested purely on spectroscopic samples of different types of galaxies, often in a limited redshift range. Additionally, some of the spectroscopic galaxy samples were simulated entirely.

\subsection{This work}
Here, we explore different kinds of mixture models to produce appropriate photo-$z$ probability distributions that naturally account for the superposition of multiple, a priori unknown classes of astrophysical objects (e.g., stars, galaxies, quasars). 
There are multiple ways to describe a distribution of such objects in photometry space that consists of e.g., magnitudes and extinction estimates (see Sect.~\ref{SS:data}) and that is also termed \textit{feature space} following the standard machine-learning terminology.

First, we use an IGMM \citep[][]{Teh2010} to separate the astrophysical objects in feature space. This approach allows the algorithm to cluster the objects based on all the available photometric information without forcing the algorithm to classify the objects in a pre-determined way. Subsequently, the structure of the photometric (feature) space defines the number of Gaussian mixture components. Whenever a spectroscopic sub-sample of different types of astrophysical objects is available, IGMMs allow to separate this sample into classes, ideally  representing each type of object. 
Secondly, we train MDNs to predict the photo-$z$ probablity distributions of objects in our data-set. To find the optimal results, we explore different MDN implementations, which all include the IGMM components and membership probabilities obtained in the first step next to the entire photomoetric (feature) space (Sect.~\ref{SS:data}).

In Section~\ref{Data and Methods}, we describe our chosen training and test data-sets as well as the IGMM and MDN implementations. The obtained accuracy of the classification along with the precision of the inferred photo-$z$s are provided in Section~\ref{Results}. In Section~\ref{Discussion} we discuss our results, shortcomings and future improvements on our photo-$z$ estimation alongside a comparison with other methods to estimate photo-$z$s from the literature.

\section{Data and Methods}
\label{Data and Methods}

To train our machine learning algorithms, we require a data-set that contains: (\textsc{i}) morphological information from publicly available object catalogs (e.g. psf \textit{vs} model magnitudes, or stellarity index), to aid the separation of stars from galaxies and quasars; (\textsc{ii}) a wide footprint of the sky, to cover regions with sufficiently different extinction; (\textsc{iii}) multi-band photometry from optical to mid-IR wavelengths, possibly including $u$-band; and (\textsc{iv}) a spectroscopic sub-sample of different types of objects (here: stars, galaxies and quasars) 

\begin{figure}
    \centering
    \includegraphics[width=0.5\textwidth]{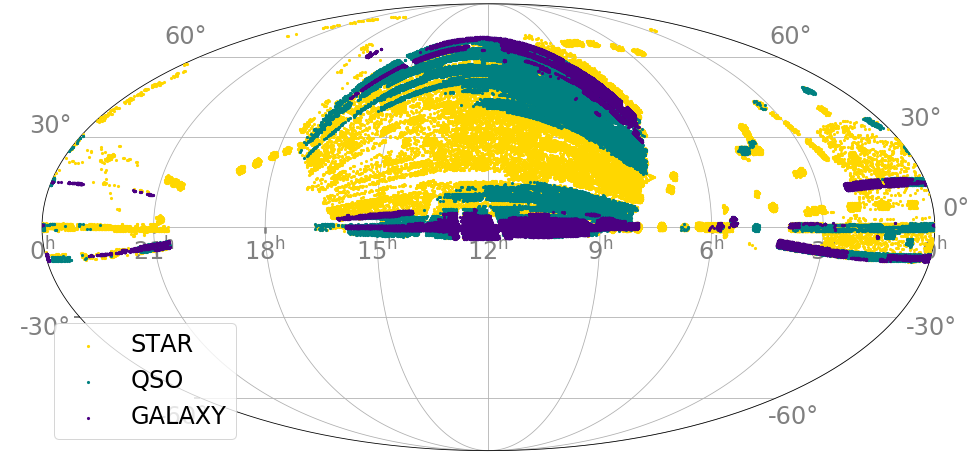}
    \caption{Spectroscopic data-set in equatorial coordinates. Data are taken from SDSS-DR15 + WISE totalling about $245\,000$ objects of which there are $86\,412$ stars (yellow), $83\,119$ galaxies (purple) and $75\,955$ quasars (green). The entire photometric data-set is a sample of about $1\,023\,000$ objects, of which $98\%$ lack {spectroscopic redshifts and classification.}
    }
    \label{fig:dataradec}
\end{figure}
\subsection{Data}
\label{SS:data}
Our photometric data-set is composed of optical \texttt{PSF} and \texttt{model} $griz$-band magnitudes including $i-$band extinction measurements from the SDSS-DR15 \citep{SDSSDR15}. We combine these SDSS magnitudes 
with \texttt{w1mpro} and \texttt{w2mpro} magnitudes (hereafter $W1,$ $W2$) from WISE \citep{Wright2010}.
We query the data in CasJobs\footnote{https://skyserver.sdss.org/casjobs/} on the \texttt{PhotoObjAll} table with a SDSS-WISE cross-match, requiring magnitude errors lower than $0.3$ mag and $i-W1<8$ mag. Adding $g-r$, $r-i$, $i-z$, $z-W1$ and $W1-W2$ colours leaves us with 22 dimensions to be used by our MDNs. However, the colours are strictly speaking redundant as they are obtained from the same, individual photometric bands. While this will introduce many null-value Eigenvectors in the IGMM, additional combinations of measurements are enabled, which will speed up the MDN computations by de-trending the magnitude-magnitude distribution.
Our spectroscopic data-set (from SDSS-DR15) includes only objects with uncertainties on their spectroscopic redshift (from the SDSS pipelines) smaller than 1\%. For only one MDN training, we added $u-$band \texttt{PSF} as well as \texttt{model} magnitudes.
Our individual data-sets are composed as follows:
\begin{itemize}
    \item Photometric data-set: $\approx2\%$ of all data have spectroscopic information. In total we have 1\,022\,731 unique sources in \texttt{PhotoObjAll} and WISE, with additional 11\,358 unique galaxies from WiggleZ \citep{2010MNRAS.401.1429D} cross-matched with \texttt{PhotoObjAll} and WISE for the IGMM.
    \item Spectroscopic data-set: 86\,412 unique stars, 83\,119 unique galaxies and 75\,955 quasars from \texttt{SpecPhoto} and WISE, for the test samples, according to the classification of their spectra by the SDSS pipelines\ref{fig:dataradec};
\end{itemize}

\subsection{Infinite Gaussian Mixture Models}
\label{SS:IGM}

In a Gaussian Mixture Model (GMM), the density distribution of objects in \textit{feature space} (equivalent to photometric space, see Sec~\ref{SS:data}) is described by a sum of Gaussian density components.
The GMM is a probabilistic model which requires that a data-set is drawn from a mixture of Gaussian density functions. Each Gaussian distribution is called a \textit{component}. As the Gaussian distributions are defined in all the dimensions of the feature space, they are characterised by a mean vector and a covariance matrix. The feature vector contains the photometric information of each astronomical source. To describe the GMM, whenever needed, we use the notation
$ \pi_{k} \mathcal{N}(x|\mu_{k}, \Sigma_{k}),$ where $k(\in\{1,...,K\})$ is the component index, $\mu_{k}$, $\Sigma_{k}$ and $\pi_{k}$ are the mean vector and the covariance matrix in feature space, and the weight of component $k$, respectively.

Since the GMM is a Bayesian method, it requires multiple sets of model parameters and hyperparameters. The model parameters (means, covariances) change across the Gaussian components, while the hyperparameters are common to all of the Gaussian components, because they describe the priors from which all Gaussian components are drawn. For the GMM, the number of Gaussian components is a fixed hyperparameter. 

The IGMM is the GMM case with an undefined number of components, which will be optimised by the model itself, depending on the photometric data-set used. In particular, the IGMM describes a mixture of Gaussian distributions on the data population with an infinite (countable) number of components, using a Dirichlet process \citep{Teh2010} to define a distribution on the component weights.

However, setting an initial number of Gaussian density components is required by the IGMM. Based on the weights that are given to each such component at the end of the model training, it is common practice to exclude the least weighted components and define the data population only by the highest-weighted components. To pursue a fully Bayesian approach, it is advisable to explore a set of model hyperparameters with different initial guesses for the number of components.
Like its finite GMM counterpart, each realisation of IGMM estimates the membership probability of each data point to each component. Appendix~\ref{app:a} provides a summary of the IGMM formalism.

For this work, we used the built-in variational IGMM package from the \texttt{scikit-learn} library for our implementations. In practice, the variational optimizer uses a truncated distribution over component weights with a fixed maximum number of components, known as stick-breaking representation \citep{ferguson1973}, with an expectation-maximization algorithm \citep{doi:10.1111/j.2517-6161.1977.tb01600.x}. To optimize the model and find the best representation of the data-set, we explore the following set of hyperparameters:
\begin{itemize}
  \item Maximum number of allowed Gaussian components: between 10 and 100, in increments of 2.
  \item Maximum number of iterations for expectation maximization performance: 2\,000.
  \item Dirichlet concentration $(\gamma)$ of each Gaussian component ($k$) on the weight distribution: $(0.01 , 0.05 , 0.0001)$ times the number of objects in the training data-set.
  \item Type of the covariance matrix for each Gaussian component: full. As per definition, each component has its own general covariance matrix.
  \item The prior on the mean distribution for each Gaussian component: median of the entries of the input vectors of the training data-set (i.e. magnitudes, extinction).
\end{itemize}

\begin{figure}
    \centering
    \includegraphics[width=0.47\textwidth]{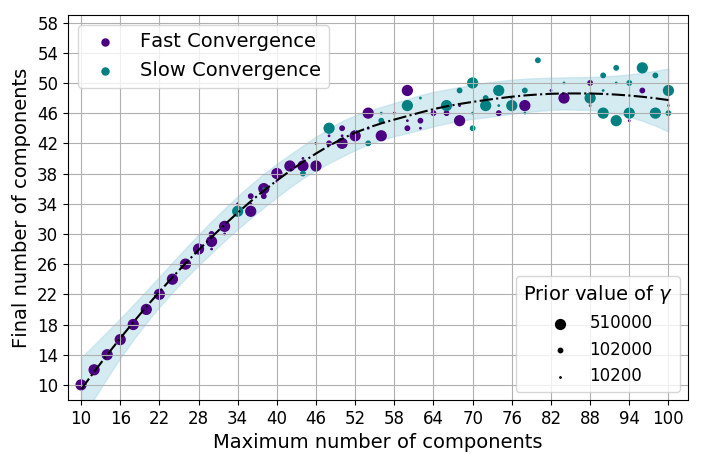}
    \caption{Maximum number of components vs. final number of components for different IGMM realisations, restricted to Gaussian components that contain at least $0.5\%$ of the photometric data. 
    Blue filled circles represent IGMM realisations that needed more than 2\,000 iterations to converge, while purple filled circles mark IGMM realisations that needed less than 2\,000 iterations. The size of the symbols scales with three different values of the prior of the Dirichlet concentration ($\gamma$).  
    The light blue shaded region represents the confidence interval of 99\% of regression estimation over the IGMM profiles by a multivariate smoothing procedure.
}
    \label{fig:ncompo_nfinal}
\end{figure}
Whenever needed, each object is assigned to the component to which its membership probability is maximal. In that case, we say that a component \textit{contains} a data-point.

The IGMM provides different possible representations of the same data-set for each set of hyperparameters: here, we are interested in finding out the optimal number of components that can adequately describe the majority of the data. We then introduce a lower threshold on the number of sources that each component contains, and drop the components which contain less than the threshold. 
The threshold is defined by considering the size of the photometric sample and the highest value that we considered for the Dirichlet $\gamma$ prior. The IGMM starts with components that contribute to 0.5\% of the size of the photometric sample, since the highest $\gamma$ value is 510\,000 (see Appendix for further details), due to our chosen ranges of hyperparameters. Therefore, we use
0.5\% of the size of the photometric data-set as the threshold. Figure~\ref{fig:ncompo_nfinal} shows that the final number of components converges to $48 \pm 4.$ The convergence indicates that the models do not need more than $48 \pm 4$ components to describe the sample. Moreover, the initial 1:1 ramp-up in the figure shows that the final number of components is the same as the maximum tolerance, and so the model cannot adequately describe the data-set; this trend breaks at about 44 components. To guide the eye, we determine a regression surface of all the IGMM profiles by a multivariate smoothing procedure\footnote{https://has2k1.github.io/scikit-misc/loess.html}. In what follows, we choose 52 components.

The first IGMM implementation was fully unsupervised, i.e. it was optimised to only describe the distribution of the objects in feature space. Subsequently, we trained different IGMMs considering additional spectroscopic information available for $\approx2\%$ of the photometric sample. In particular, these \textit{partially supervised} implementations are trained using the entire photometric feature space including either (\textsc{i}) spectroscopic classifications or (\textsc{ii}) spectroscopic redshifts or (\textsc{iii}) spectroscopic classifications and redshifts. Since the objects with additional spectroscopic information are a small part of the photometric training sample ($\approx2\%$), the implementations ensure that the SDSS spectroscopic pre-selection does not bias the IGMM over the entire photometric sample.
Finally, we calculate the membership probabilities to the 52 components for each object in the spectroscopic data-set ($\approx2.45\times10^{5}$ objects) from the optimised IGMM. This allows us to assign each object from the spectroscopic sample to one component. Thereafter, we label each of the IGMM components based on the percentage of spectroscopic classes that it contains. 

\begin{figure*}
    \centering
    \includegraphics[width=0.99\textwidth]{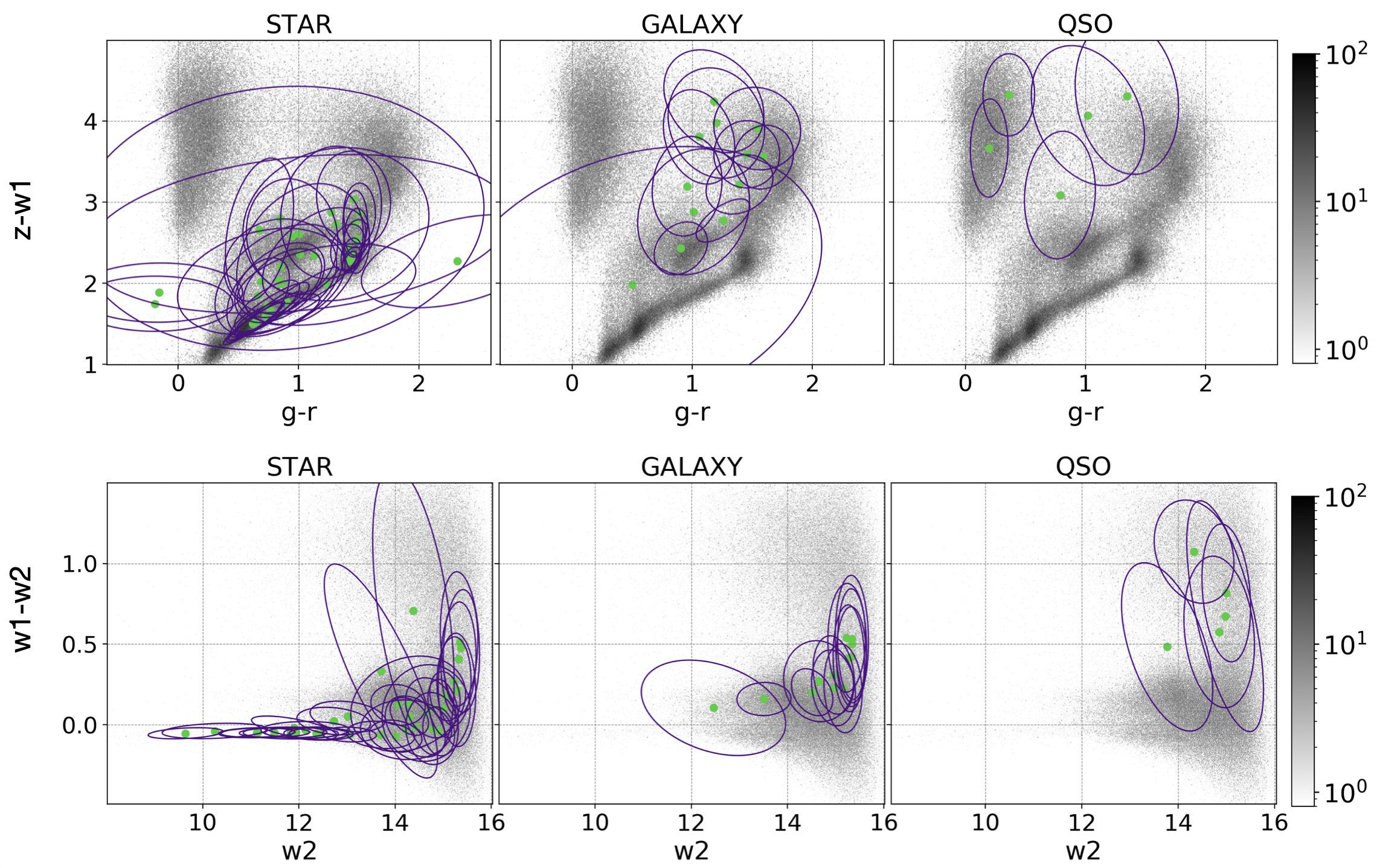}
\caption{
    Colour-colour and colour-magnitude diagrams. Shown are $g-r$ vs $z-W1$ colour-colour diagrams (upper panel) and $W2$ vs $W1-W2$ colour-magnitude diagrams (bottom panel) for a populations of objects from the spectroscopic data-set such as stars (left column), galaxies (middle column) and quasars (right column). The purple contours correspond to the 68-th percentile of each Gaussian IGMM component. The green filled circles correspond to the means $\boldsymbol{\mu}_{k}$ of the Gaussian components. The grey scale indicates the abundance of the sources in each diagram. 
}
    \label{fig:colmag}
\end{figure*}

Figure~\ref{fig:colmag} shows the population of objects from the spectroscopic data-set and their corresponding IGMM components in $g-r$ vs. $z-w1$ (upper panel) and $w2$ vs. $w1-w2$ (bottom panel) colour-colour and colour-magnitude diagrams. Each row from left to right shows the assigned components to stars, galaxies and quasars in the respective panels.

\subsection{Mixture Density Networks}
\label{ss:Mixture Density Network}
MDNs are a form of ANNs, which are capable of arbitrarily accurate approximation to a function and its derivatives based on the \textit{Universal Approximation Theorem} \citep{Hornik1991ApproximationCO}. ANNs can be used for regression or classification purposes. ANNs are structured in layers of neurons, where each neuron receives an input vector from the previous layer, and outputs a non-linear function of it that is passed on to the next layer. In MDNs, the aim is to approximate a distribution in the product space of input vectors of the individual sources ($\mathbf{f}_{i}$) and target values (e.g., $z_{s,i}$) as a superposition of different components. MDNs \citep{astonpr373} are trained to optimize the log-likelihood

\begin{equation}
    \log\mathcal{L}\ =\ \sum\limits_{i=1}^{N}\log\left( \sum_{k=1}^{N_{c}}\hat{p}_{k}(\mathbf{f}_{i})\mathcal{N}(z_{s,i}|m_{k}(\mathbf{f}_{i}),s_{k}(\mathbf{f}_{i}))\right)
\end{equation}
by approximating the averages $m_{k}(\mathbf{f}),$ amplitudes $\hat{p}_{k}(\mathbf{f})$ and widths $s_{k}(\mathbf{f})$. Here, $N$ is the number of objects in the spectroscopic data-set, while $N_{c}$ denotes the number of output components (or \textit{branches}) of the MDN.

Due to the limited information provided by the photometric space, a source of a specific spectroscopic class and low redshift can be confused with a different spectroscopic class and high redshift. Therefore, by providing distributions over a full range of redshifts, MDNs can cope with the fact that colours are not necessarily monotonic with redshift (as is the case e.g. in quasars). In order to avoid confusing MDN components with IGMM components, here we call MDN components \textit{branches}.

For the sake of reproducibility, we use a publicly available MDN wrapper around the \texttt{keras} ANN module\footnote{https://github.com/cpmpercussion/keras-mdn-layer} and a simple MDN architecture. 
The MDN input layer contains the same photometric features (see \ref{SS:data}) along with the membership probabilities of the IGMM, which carry additional information of the object classes (stars, galaxies and quasars). The dimension of the MDN input space is 74, of which 52 are the IGMM membership probabilities and 22 are the feature-space entries. The output layer of the MDN is defined by three neurons for each branch: the average redshift on the branch, the width of the branch and the membership probability of the source to the branch. The MDN is fully connected, i.e. the neurons in one layer are connected to all of the neurons in the next layer. Due to the fact that the MDN input contains the IGMM membership probabilities, after MDN hyperparameter optimization, we train one MDN for each of the four IGMM implementations as described in previous sections.

\begin{figure}
    \centering
    \includegraphics[width=0.4\textwidth]{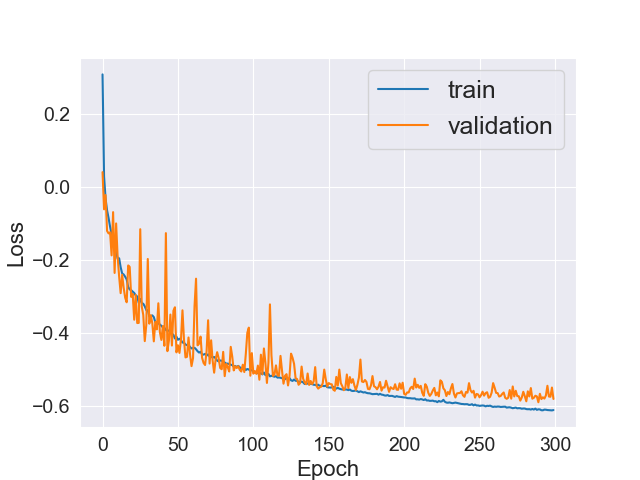}
    \caption{MDN Loss ($-\log(\mathcal{L})/N$) as a function of epoch. The loss obtained during the MDN training and validation are shown by blue and orange lines, respectively.}
  
    \label{fig:loss-mdn}
\end{figure}
\subsubsection{Hyperparameter selection and tuning}

We randomly split the entire spectroscopic data-set (\ref{SS:data}) and use 80\% for training and 20\% for validation of the MDN. In order to optimize the MDN, we explored the following hyperparameters:
\begin{itemize}
    \item Number of hidden neurons in the dense layer: 3, 7, 10, 74, 100, 156, 222, 300, 400, 500, 528, 600, 740
    \item Number of hidden layers: 0,1,2,3
    \item Number of MDN branches: 10, 52, 56, 100, 300.
    \item Activation function for dense layer: standard rectified linear unit \citep[ReLU,][]{10.5555/3104322.3104425} and parametric rectified linear unit \citep[PReLU,][]{2015arXiv150201852H}
    \item Learning rate: $10^{-6},$ $10^{-5},$ $10^{-4},$  $10^{-3}$ 
\end{itemize}
To mitigate local minima of the loss function, we used \texttt{ADAM} as optimizer and batch learning with 64 objects per epoch.

By comparing the training and validation loss of MDNs with the previously defined set of hyperparameters, the resulting optimal set of hyperparameters contains:
\begin{itemize}
    \item Hidden neurons in the dense layer: 528
    \item Number of MDN branches: 10
    \item Activation function for dense layer: PReLU
    \item $10^{-4}$ learning rate
\end{itemize}
Figure~\ref{fig:loss-mdn} shows the loss function, $-\log(\mathcal{L})/N$, for the training and validation data-set, for the MDN optimisation for which membership probabilities are obtained from the partially supervised IGMM realisation that also considers the spectroscopic classes. As Figure~\ref{fig:loss-mdn} shows, the learning curve flattens roughly around 300 epochs. To mitigate over-fitting, we concluded that 300 epochs are sufficient to train the model.
Additionally to training MDNs with the redshifts as targets, we tested $\log(z_{s})$ as a target and it led to an improvement in the $z_{p}$ estimation. 

\begin{figure}
    \centering
    \includegraphics[width=0.4\textwidth]{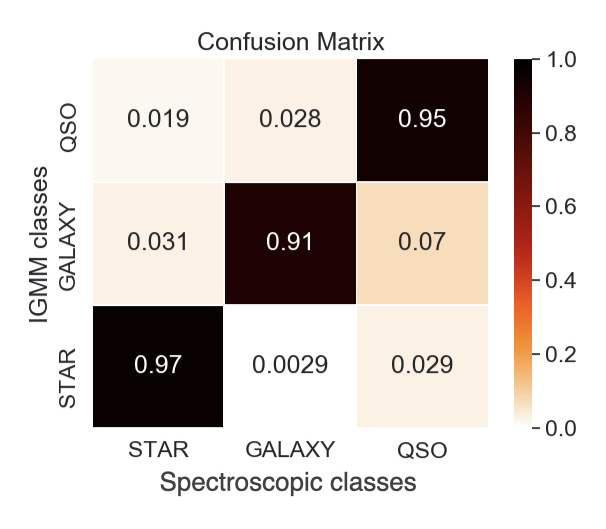}
    \caption{IGMM confusion matrix. The spectroscopic classifications are shown against the IGMM classes of the spectroscopic data-set.
    }
    \label{fig:confusion matrix for GMM}
\end{figure}

\begin{table}[h!t]
\caption{Percentage of objects from each spectroscopic class (stars, galaxies, quasars) within each IGMM component. The components highlighted in red lie between different spectroscopic class regions in photometric feature space, and can reduce the classification accuracy.}

\label{table:per-IGMM-blob}      
\centering          
\begin{tabular}{p{1.6cm}|p{1.5cm}|p{1.5cm}|p{1.5cm}} 
    
 IGMM  & Stars & Galaxies & Quasars\\
 components & & &
 
  \\
  \hline\hline
  \textcolor{red}{1} & 85.42 & 0.27 & 14.31
  \\
  \hline
    2 & 99.98 & 0 & 0.02
  \\
  \hline
  3 & 97.46 & 0.06 & 2.48
  \\
  \hline
  4 & 100 & 0 & 0
  \\
  \hline
  5 & 1.57 & 0.54 & 97.88
  \\
  \hline
  6 & 99.86 & 0.05 & 0.1
  \\
  \hline
  \textcolor{red}{7} & 3.7 & 86.06 & 10.24
  \\
  \hline
  8 & 100 & 0 & 0
  \\
  \hline
  9 & 97.45 & 0.05 & 2.5
  \\
  \hline
  \textcolor{red}{10} & 8.94 & 71.67 & 19.39
  \\
  \hline
  11 & 1.97 & 90.16 & 7.87
  \\
  \hline
  \textcolor{red}{12} & 6.95 & 52.25 & 40.80
  \\
  \hline
  13 & 99.6 & 0 & 0.4
  \\
  \hline
  14 & 100 & 0 & 0
  \\
  \hline
  \textcolor{red}{15} & 42.38 & 43.22 & 14.41
  \\
  \hline
  \textcolor{red}{16} & 55.39 & 0.43 & 44.18
  \\
  \hline
  17 & 99.93 & 0.01 & 0.06
  
  \\
  \hline
  18 & 96.75 & 2.48 & 0.77
  \\
  \hline
  \textcolor{red}{19} & 6.58 & 36.44 & 56.98
  \\
  \hline
  20 & 99.89 & 0 & 0.11
  \\
  \hline
  21 & 1.14 & 94.51 & 4.35
  \\
  \hline
  22 & 98.02 & 0.07 & 1.90
  \\
  \hline
  23 & 99.94 & 0 & 0.06
  \\
  \hline
  24 & 3.69 & 89.54 & 6.77
  \\
  \hline
  25 & 100 & 0 & 0
  \\
  \hline
  26 & 99.94 & 0.01 & 0.05
  \\
  \hline
  27 & 97.48 & 0.47 & 2.05
  \\
  \hline
  28 & 100 & 0 & 0
  \\
  \hline
  \textcolor{red}{29} & 12.31 & 20.04 & 67.65
  \\
  \hline
  30 & 100 & 0 & 0
  \\
  \hline
  31 & 1.02 & 96.60 & 2.38
  \\
  \hline
  \textcolor{red}{32} & 11.13 & 35.58 & 53.28
  \\
  \hline
  33 & 99.96 & 0.02 & 0.02
  \\
  \hline
  34 & 99.71 & 0 & 0.29
  \\
  \hline
  35 & 100 & 0 & 0
  \\
  \hline
  36 & 99.8 & 0.1 & 0.1
  \\
  \hline
  \textcolor{red}{37} & 34.23 & 42.05 & 23.72
  \\
  \hline
  38 & 100 & 0 & 0
  \\
  \hline
  \textcolor{red}{39} & 8.43 & 51.74 & 39.83
  \\
  \hline
  40 & 99.91 & 0 & 0.09
  \\
  \hline
  41 & 99.51 & 0.04 & 0.45
  \\
  \hline
  42 & 100 & 0 & 0
  \\
  \hline
  \textcolor{red}{43} & 4.43 & 88.61 & 6.97
  \\
  \hline
  44 & 0.56 & 98.18 & 1.25
  \\
  \hline
  45 & 90.3 & 0.83 & 8.87
  \\
  \hline
  \textcolor{red}{46} & 79.57 & 1.22 & 19.21
  \\
  \hline
  \textcolor{red}{47} & 2.87 & 65.41 & 31.72
  \\
  \hline
  48 & 0.73 & 0.05 & 99.21
  \\
  \hline
  49 & 100 & 0 & 0
  \\
  \hline
  \textcolor{red}{50} & 60.24 & 0.74 & 39.02
  \\
  \hline
  \textcolor{red}{51} & 44.52 & 26.33 & 29.15
  \\
  \hline
  52 & 95.64 & 0.04 & 4.32
  \\
  \hline

\end{tabular}
\end{table}

\section{Results}
\label{Results}

We trained an IGMM on the photometric data-set (see sect.~\ref{SS:data}), using the optimal hyperparameters (sect.~\ref{SS:IGM}). Thereafter, we linked IGMM components to the three spectroscopic classes using a spectroscopic data-set (\ref{SS:data}).
Finally, we implemented MDNs on the spectroscopic data-set using photometric features and membership probabilities from the IGMM to estimate the conditional probability distribution $p(z_{p}|\mathbf{f})$ of photo-$z$ values from the photometric inputs. In this section, we describe the evaluation methods and the resulting classification and photo-$z$ estimations.

\subsection{Classification}

With our mixture models we address the common problem of cross-contamination among different classes of objects due to the a priori unknown underlying spectral-energy distribution.
In the IGMM realisations, each object can belong to each of the components with a probability $p_{i,k}=w_{k}\mathcal{N}(\mathbf{f}_{i}|\boldsymbol{\mu}_{k},\boldsymbol{\Sigma}_{k})/\sum_{l}(w_{l}\mathcal{N}(\mathbf{f}_{i}|\boldsymbol{\mu}_{l},\boldsymbol{\Sigma}_{l}))$, which we will denote by \textit{membership probabilities} in the following. As we introduced above (end of Sect.~\ref{SS:IGM}), the simplest way to assign an object (with feature vector $\mathbf{f}_{i}$) to a component is to consider the component index $\hat{k}$ for which $p_{i,\hat{k}}$ is maximised. 

To parameterize the accuracy of the classification, we consider the usual quantification of true/false positives and true/false negatives \citep[e.g.][]{journals/prl/Fawcett06}, and build a \textit{confusion matrix} to quantify the rate of correct classifications.
Figure~\ref{fig:confusion matrix for GMM} shows the confusion matrix of the GMM-based classification for the spectroscopic data-set. The true positive rates\footnote{Defined as: TP/(TP+FN).} for stars, galaxies and quasars are 0.97, 0.91 and 0.95, respectively. False positive rates for stars that are true galaxies and quasars are 0.0029 and 0.029. False negative rates for stars that are assigned to galaxies and quasars are 0.031 and 0.019 of all stars, respectively.
The accuracy\footnote{Defined as: (TP+TN)/(TP+TN+FN+FP).} is $\approx 94 \%.$
This means that the IGMM part of our mixture models can clean an extragalactic sample from most of stellar contaminants, and broadly separate galaxies from AGN-dominated objects.

Figure~\ref{fig:colmag} demonstrates that the IGMM recognizes the main behaviours of stars, galaxies and quasars in colour space and also identifies sub-classes that are not highly represented in the spectroscopic sample, such as white dwarfs and brown dwarfs. On the other hand, some components happen to lie in regions of the colour-magnitude-extinction space that are not dominated by only one sub-class.
The overlap between different object classes in photometry can affect the classification performance and the output of the classification that is then used by the MDN regression. The components corresponding to regions of overlap between different classes are discussed below.

\begin{figure*}
    \centering
    \includegraphics[width=0.99\textwidth]{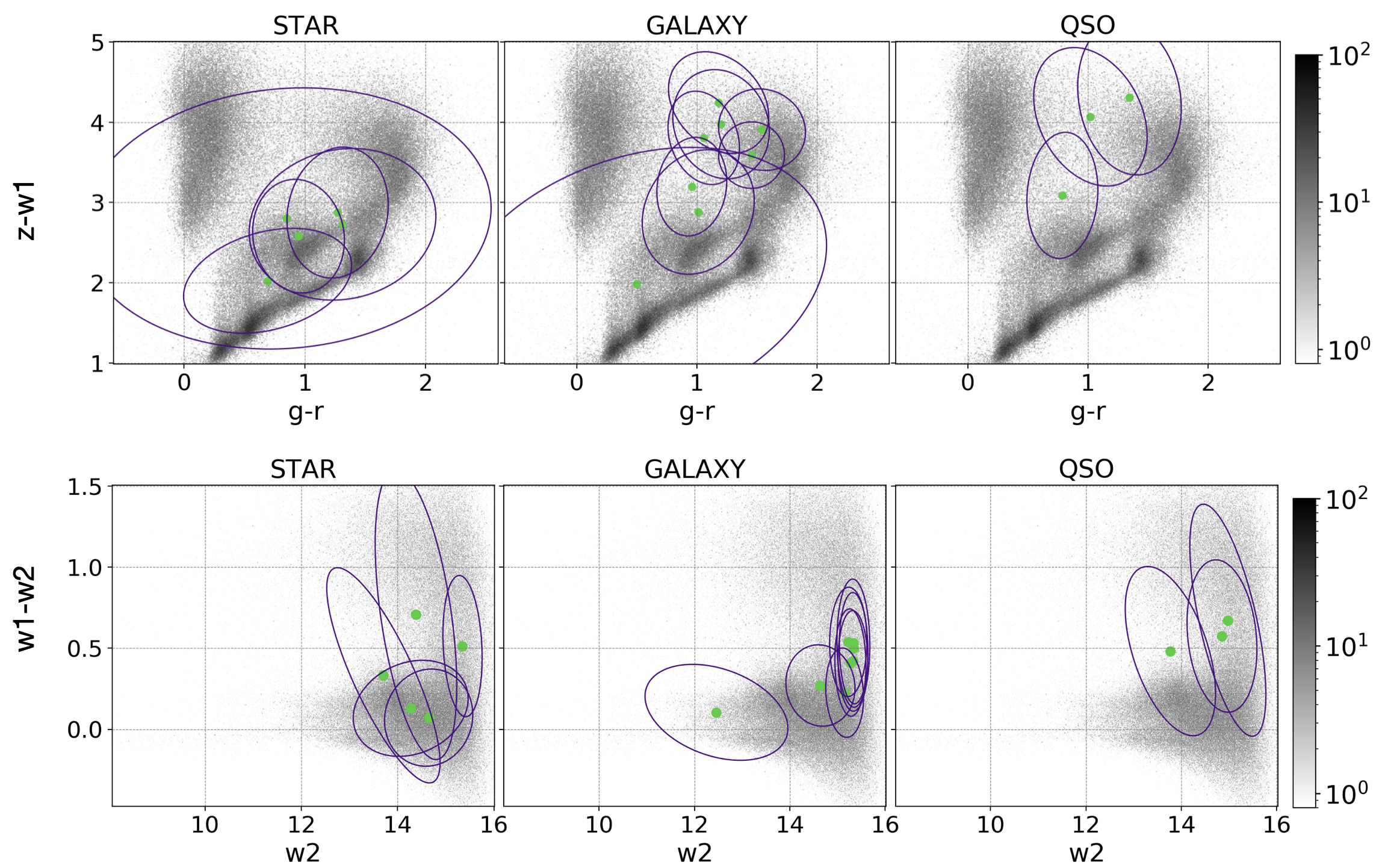}
 \caption{Colour-colour and colour-magnitude diagrams. Shown are $g-r$ vs $z-W1$ colour-colour diagrams (upper panel) and $W2$ vs $W1-W2$ colour-magnitude diagrams (bottom panel) for objects from the spectroscopic data-set of the three spectroscopic classes such as stars (left column), galaxies (middle column) and quasars (right column). The purple contours correspond to the 68-th percentile of the \textit{problematic} Gaussian components of the IGMM that are not dominated by objects of just one spectroscopic class. The green filled circles correspond to the means $\boldsymbol{\mu}_{k}$ of these components. The grey scale indicates the number of sources in each diagram.
     }
    \label{fig:problematic_blobs}
\end{figure*}
\subsubsection{Problematic colour-magnitude-extinction regions and the corresponding IGMM components}

Approximately $30\%$ of IGMM components that cover $\approx 15\%$ of the spectroscopic data-set, marked in red in Table~\ref{table:per-IGMM-blob}, contain a non-negligible fraction of objects from more than one of the three main classes.
Figure~\ref{fig:problematic_blobs} shows their position in the same colour-colour and colour-magnitude diagrams as Figure~\ref{fig:colmag}. We will address these components as `problematic components'.

As expected, the problematic components lie at the faint end
(with higher magnitude uncertainties in WISE), or in intermediate regions of the colour space between AGN-dominated and galaxy-dominated systems.
Additionally, the SDSS spectroscopic classification of some objects is ambiguous and for some cases the automatic classification (by the SDSS spectral pipelines) is either erroneous or has multiple incompatible entries\footnote{E.g. for \texttt{OBJID=}1691188859137714176 from SDSS-DR15}. These issues occur more frequent for fainter objects which have spectra with low signal-to-noise ratio\footnote{E.g. for \texttt{OBJID=}743142903307593728 from SDSS-DR15}. However, since most of the objects are clustered in three main classes which are correctly identified by the IGMM components, uncertain spectroscopic labels are not a significant problem for our calculations.

\begin{figure*}[t]
    \centering
    \includegraphics[width=0.49\textwidth]{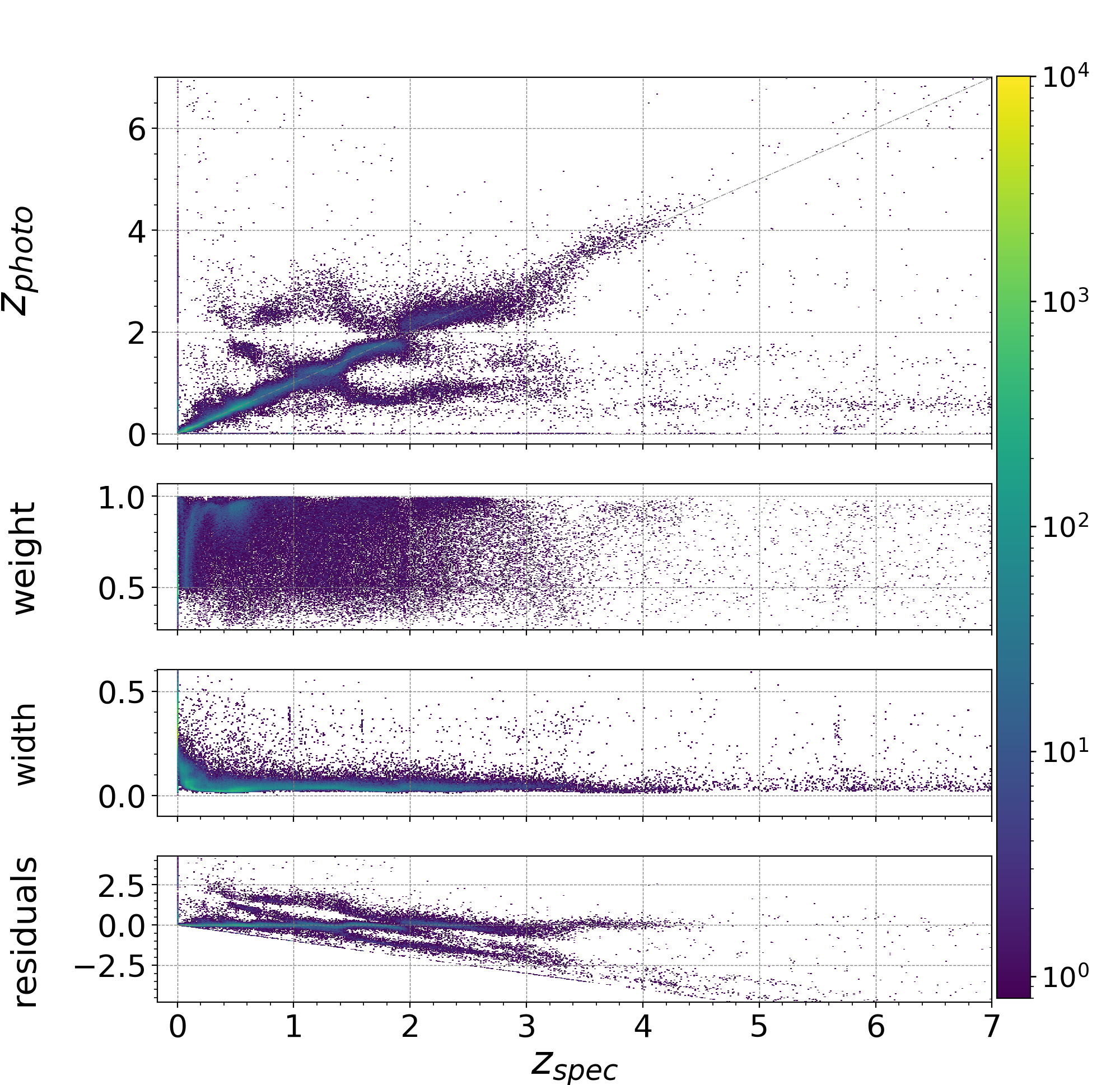}
    \includegraphics[width=0.49\textwidth]{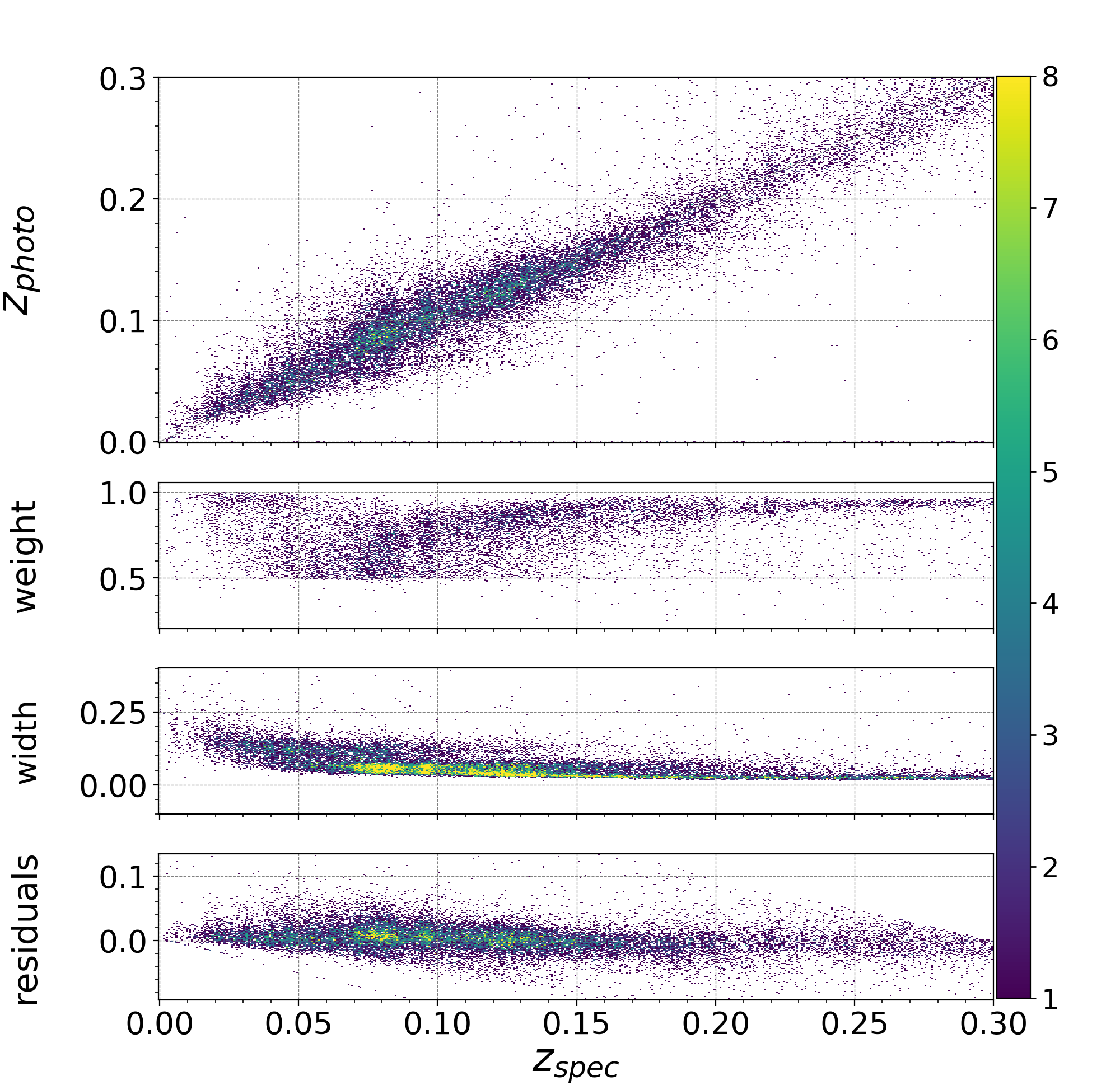}\\
    \includegraphics[width=0.49\textwidth]{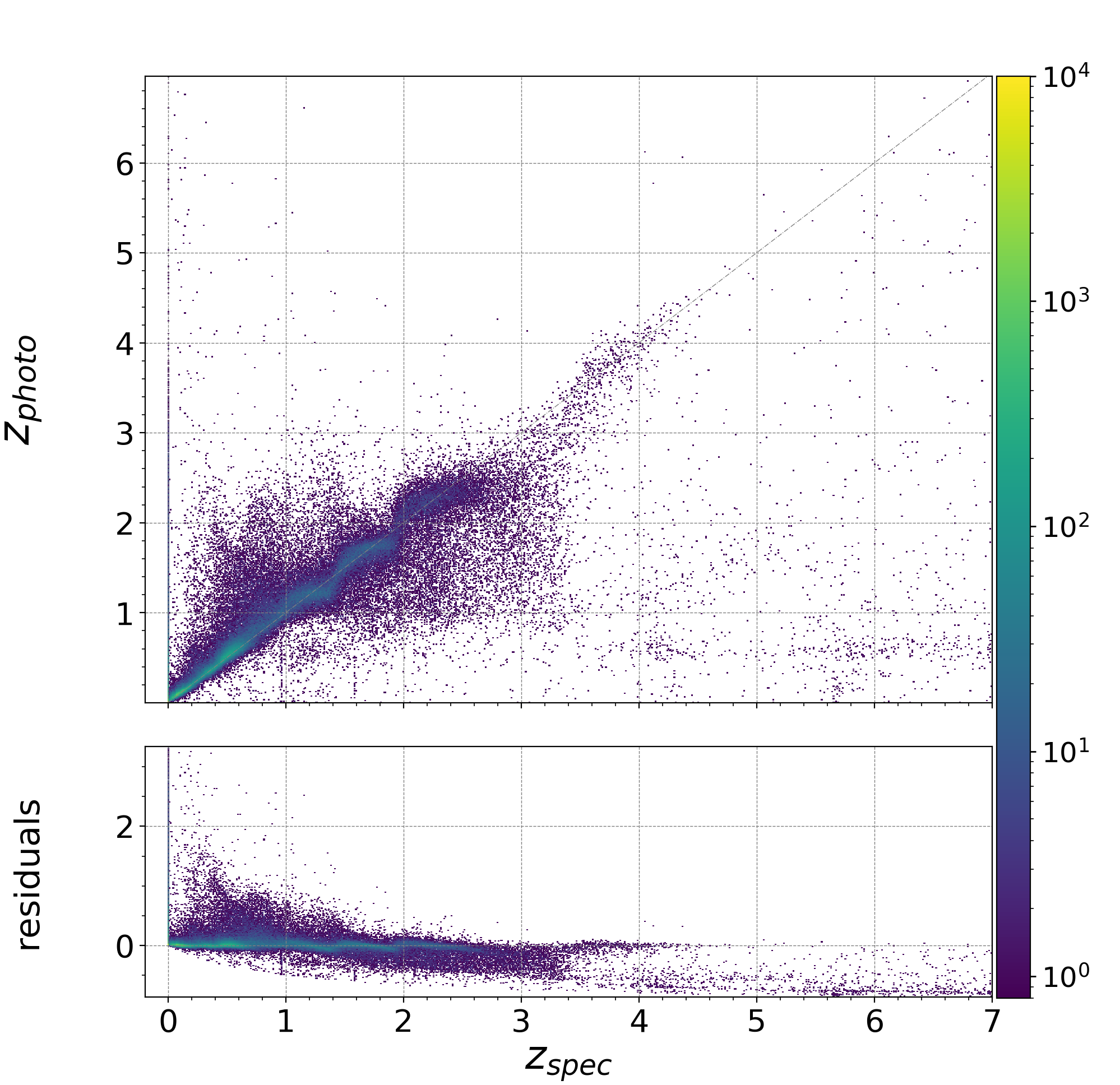}
    \includegraphics[width=0.49\textwidth]{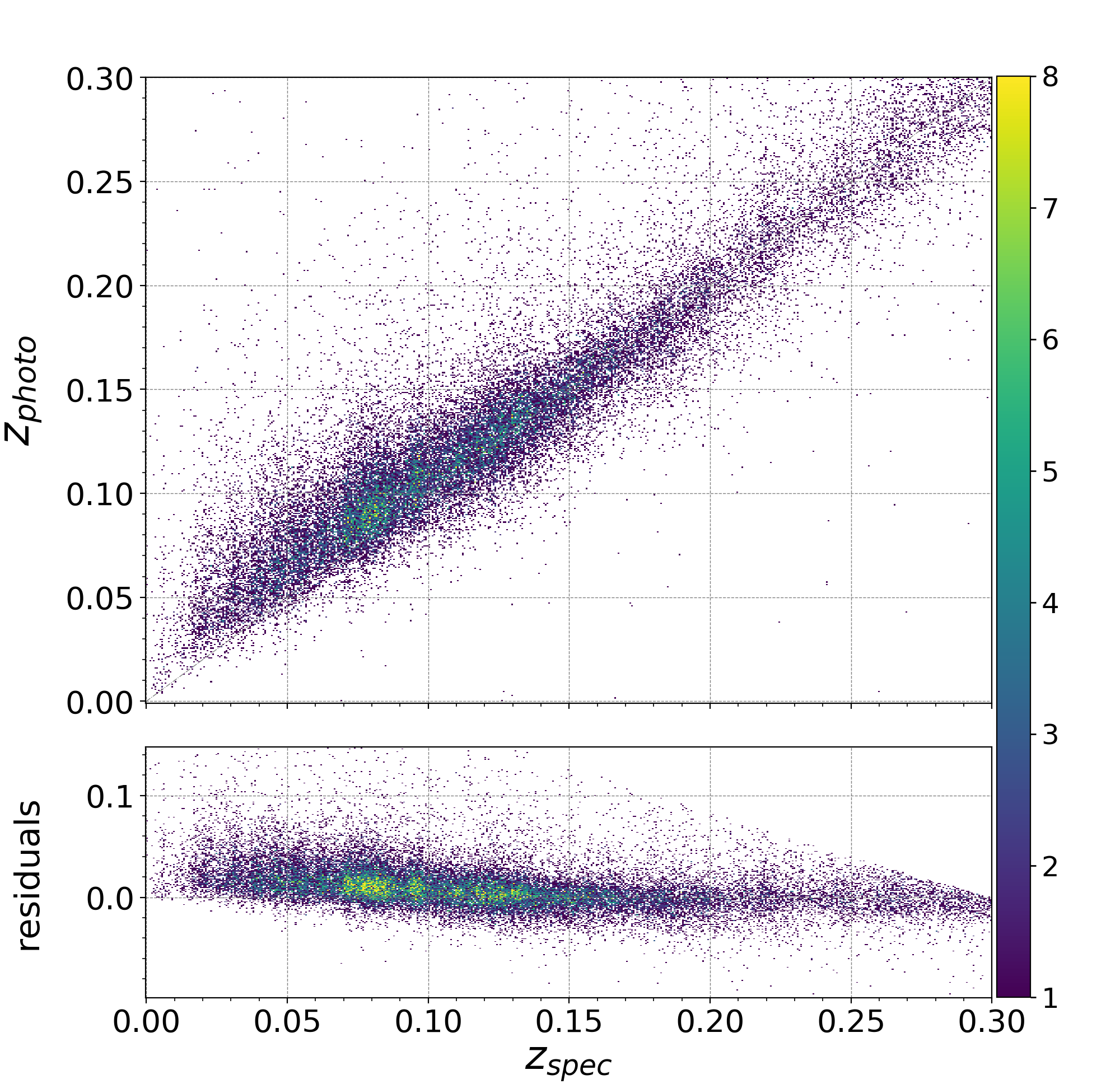}
    
    \caption{Comparison of spectroscopic vs. IGMM photometric redshifts.
    The photometric redshifts are taken from the partially supervised `spec. class' IGMM implementation (as described in Sec.~\ref{SS:IGM}). The colour-scales indicate the number of objects.
    \textit{Top panels:} The predicted photometric redshifts that correspond to the branches with the highest weights. The single panels show the weights, dispersions (denoted by "width") and residuals from top to bottom.  
    \textit{Bottom panels:} The mean photometric redshifts of the predicted redshifts over all  branches with respect to their weights. The lower panel shows the residuals.
    \textit{Left panels:} Include all classes with $z_{spec}<7$.
    \textit{Right panels:} Include all galaxies with $z_{spec}<0.3$.}
    \label{fig:Photoz-specz}
\end{figure*}

\begin{table*}
\caption{MDN performance evaluation, without any clipping for the average and rms, without any threshold on branch membership probabilities.
}
\label{table:performance}      
\centering     
\begin{tabular}{p{2.1cm}|p{1.9cm}|p{1cm}|p{1.1cm}|p{1.6cm}|p{1.1cm}|p{1.2cm}|p{1.7cm}|p{1.2cm}|p{1.2cm}} 
  IGMM 
  & photometry 
    & $\langle\Delta z\rangle$
    & rms($\Delta z$) 
    & $3\sigma$ outliers
    & $\langle\Delta z\rangle,$ 
    &  rms($\Delta z$),
    & $3\sigma$ outliers, 
    & rms($\Delta z$), 
    & rms($\Delta z$),\\
    implementation
    &
    &(all)
    &(all)
    &(all)
    &range1\tablefootmark{a}
    & range1\tablefootmark{a}
    &range1 \tablefootmark{a}
    &range2\tablefootmark{b}
    &range3\tablefootmark{c}
  \\
  \hline\hline 
  Fully unsup. & $griz,$ $W1,$ $W2$ & 0.0152 & 0.2174 & 3.08\% & 0.0007 & 0.0177 & 0.28\% & 0.0988 & 0.0945 \\
  \hline 
  spec.~class  & $griz,$ $W1,$ $W2$  & 0.0111 &0.2069 & 1.31\%&0.0006 & 0.0167& 0.41\% & 0.0822 & 0.0783\\
  \hline
  spec.~class\tablefootmark{d}  & $griz,$ $W1,$ $W2$ & 0.0356 & 0.2300 & 1.35\% & 0.0110 & 0.0260 & 0.71\% & 0.0953 & 0.0903 \\ 
  \hline 
  redshift ($z_{s}$) & $griz,$ $W1,$ $W2$  & 0.0176 & 0.2131& 3.21\% & -0.0009 & 0.0174 & 0.38\% & 0.0896 & 0.0873 \\
  \hline 
  spec.~class, $z_{s}$ & $griz,$ $W1,$ $W2$  & 0.0047 & 0.1990& 2.66\%&  0.0036 & 0.0181& 0.57\% & 0.0675 & 0.0664  \\
  \hline 
  spec.~class & $ugriz,$ $W1,$ $W2$  & 0.0135 & 0.1592 & 1.62\% & 0.0007 & 0.0160 & 0.23\% & 0.0601 & 0.0611\\
  \hline

\end{tabular}
\tablefoot{
{Spectroscopic sample for all IGMM implementations containing stars, galaxies and quasars.}\\
\tablefoottext{a}{Restricted to galaxies with $z_{s}<0.3$; } 
\tablefoottext{b}{Restricted to galaxies with $z_{s}<0.4$; }
\tablefoottext{c}{Restricted to galaxies with $z_{s}<0.5$ ;}
\tablefoottext{d}{Expectation value.}
}
\end{table*}
\begin{table*}
\caption{MDN performance evaluation exclusively for sources with MDN branch $weight_{max} > 0.8,$ without any clipping for the average and rms.
}
\label{table:performance-8}   
\centering 
\begin{tabular}{p{2.1cm}|p{1.9cm}|p{1cm}|p{1.1cm}|p{1.6cm}|p{1.1cm}|p{1.2cm}|p{1.7cm}|p{1.2cm}|p{1.2cm}} 
  IGMM 
  & photometry 
  & $\langle\Delta z\rangle$
  & rms($\Delta z$)  
  & $3\sigma$ outliers
  & $\langle\Delta z\rangle,$
  &  rms($\Delta z$),
  & $3\sigma$ outliers,
  & rms($\Delta z$),
  &rms($\Delta z$),\\
  
  implementation
  &
  &(all)
  &(all)
  &(all)
  &range1\tablefootmark{a}
  &range1\tablefootmark{a}
  &range1\tablefootmark{a}
  &range2\tablefootmark{b}
  &range3\tablefootmark{c}
  \\
  \hline\hline 
  Fully unsup. & $griz,$ $W1,$ $W2$ & 0.0032 &0.1165 &1.00\% & 0.0007 & 0.0177 & 0.60\% & 0.0350 & 0.0360\\
  \hline 
  spec.~class & $griz,$ $W1,$ $W2$  & 0.0031 & 0.1244 & 0.93\% &  0.0006 & 0.0167 & 0.83\% & 0.0405 & 0.0391 \\
  \hline 
  redshift ($z_{s}$) & $griz,$ $W1,$ $W2$  & 0.0035 & 0.1076 & 0.79\% &  -0.0009 & 0.0174 & 0.52\% & 0.0299 & 0.0331  \\
  \hline 
  spec.~class, $z_{s}$ & $griz,$ $W1,$ $W2$  & -0.0048 & 0.1170 & 1.02\% & 0.0036 & 0.0036 & 1.02\% &0.0337 & 0.0314\\
  \hline 
  spec.~class & $ugriz,$ $W1,$ $W2$  & 0.0043 &0.0934 & 0.66\% &  0.0007 & 0.0160 & 0.92\% & 0.0334 & 0.0341 \\
  \hline

\end{tabular}
\tablefoot{ 
\tablefoottext{a}{Restricted to galaxies with $z_{s}<0.3$; } 
\tablefoottext{b}{Restricted to galaxies with $z_{s}<0.4$; } 
\tablefoottext{c}{Restricted to galaxies with $z_{s}<0.5$ .}
}
\end{table*}
\subsection{Photometric redshifts}
Here we discuss different metrics employed to evaluate the performance of our methods used to determine photometric redshifts. Most metrics are based on commonly used statistical methods as outlined: 
\begin{itemize}
\item Prediction \textit{bias}: defined as the mean of weighted residuals, $\Delta z = (z_{p} - z_{s})/(1+ z_{s})$ as defined in \cite{2000ApJ...538...29C}
\item Root-mean-square of the weighted residuals: $\mathrm{rms}(\Delta z)$
\item Fraction of outliers: defined as the number of objects with $ 3 \times \mathrm{rms}(\Delta z) < \Delta z $
\end{itemize}

For all methods, we excluded objects with spectroscopic redshift errors $\delta z_{s} > 0.01 \times (1+z_{s}).$
For each source, the MDN determines a full photo-$z$ distribution, which is a superposition of all branches, each with a membership probability, average, and dispersion. If one so-called \textit{point estimate} is needed, there are at least two options to compute it.
One option is the expectation value
\begin{equation}
    \mathbb{E}(z_{p,i}|\mathbf{f}_{i})=\frac{\sum_{k}\mu_{k}(\mathbf{f}_{i})\hat{p}_{k}\mathcal{N}_{k}}{\sum_{k}\hat{p}_{k}\mathcal{N}_{k}}\ .
\end{equation}
Another, common option is the maximum-a-posteriori value, i.e. the peak $\mu_{r}(\mathbf{f}_{i})$ of the branch that gives the maximum membership probability(amplitude). 
of a given object.
We choose to compute both values and obtain a higher accuracy for the maximum-a-posteriori value than for the expectation value. 

\begin{figure*}[h!t]
    \centering
    \includegraphics[width=0.43\textwidth]{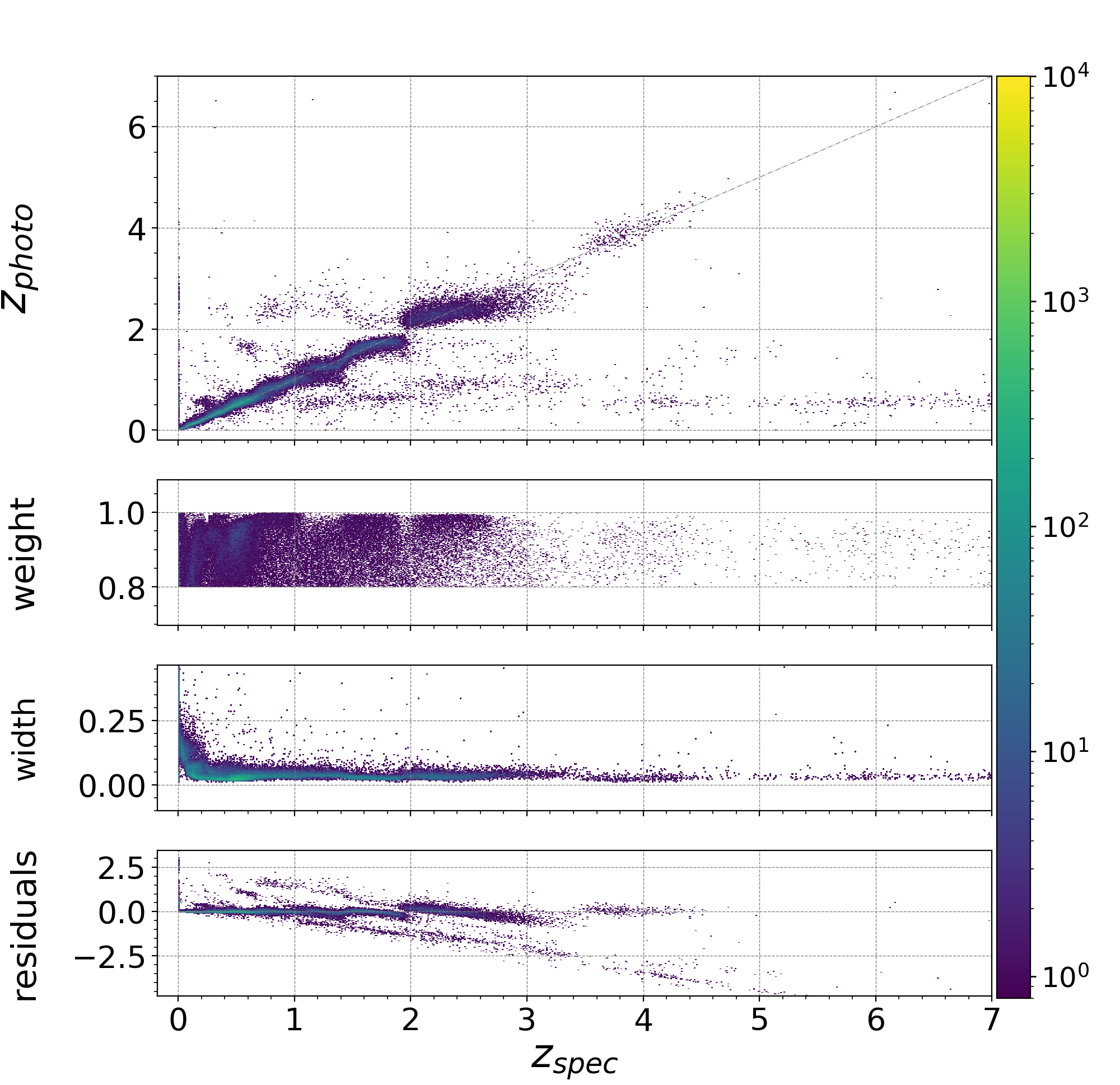}
    \includegraphics[width=0.43\textwidth]{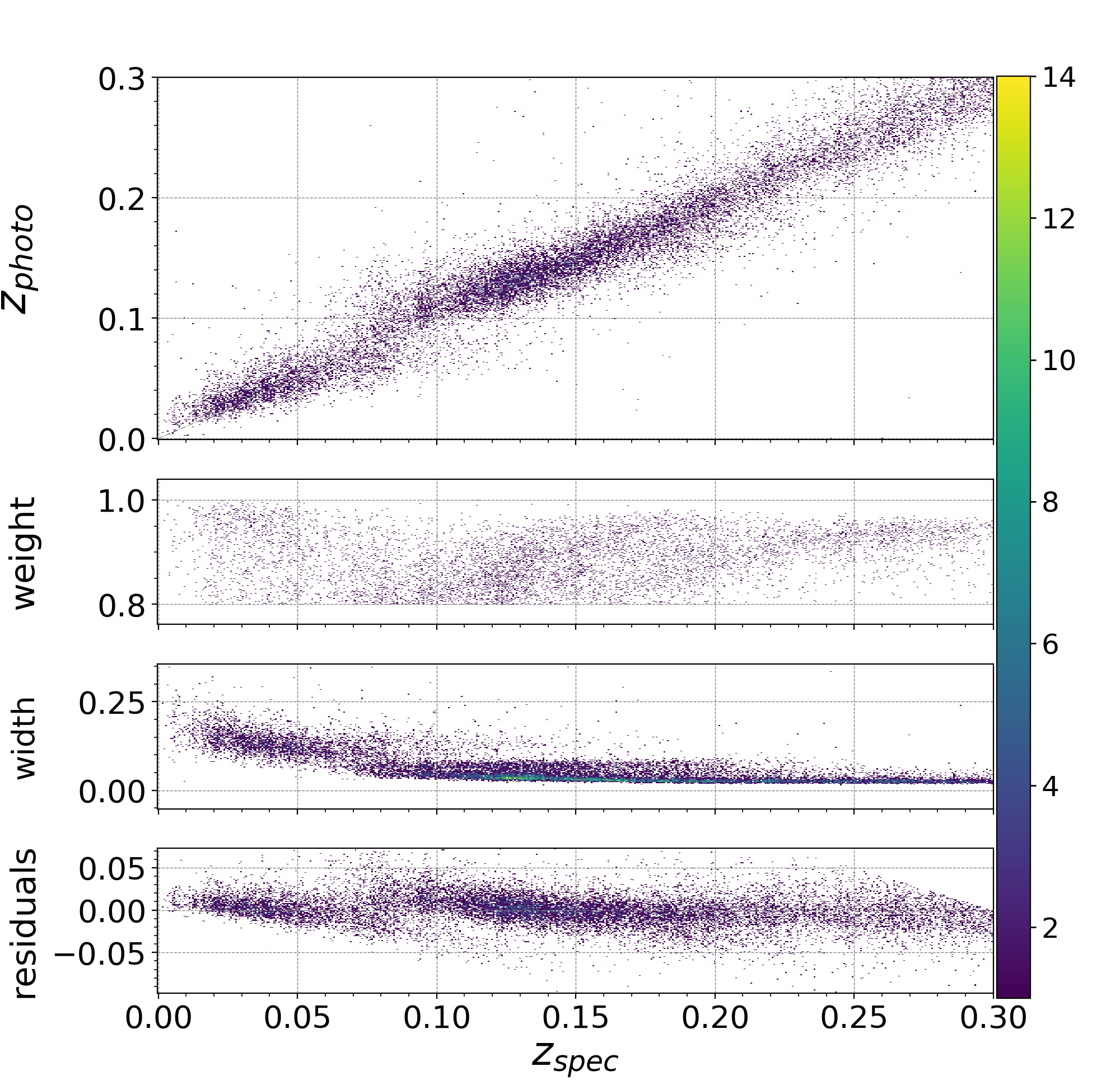}\\
    \includegraphics[width=0.43\textwidth]{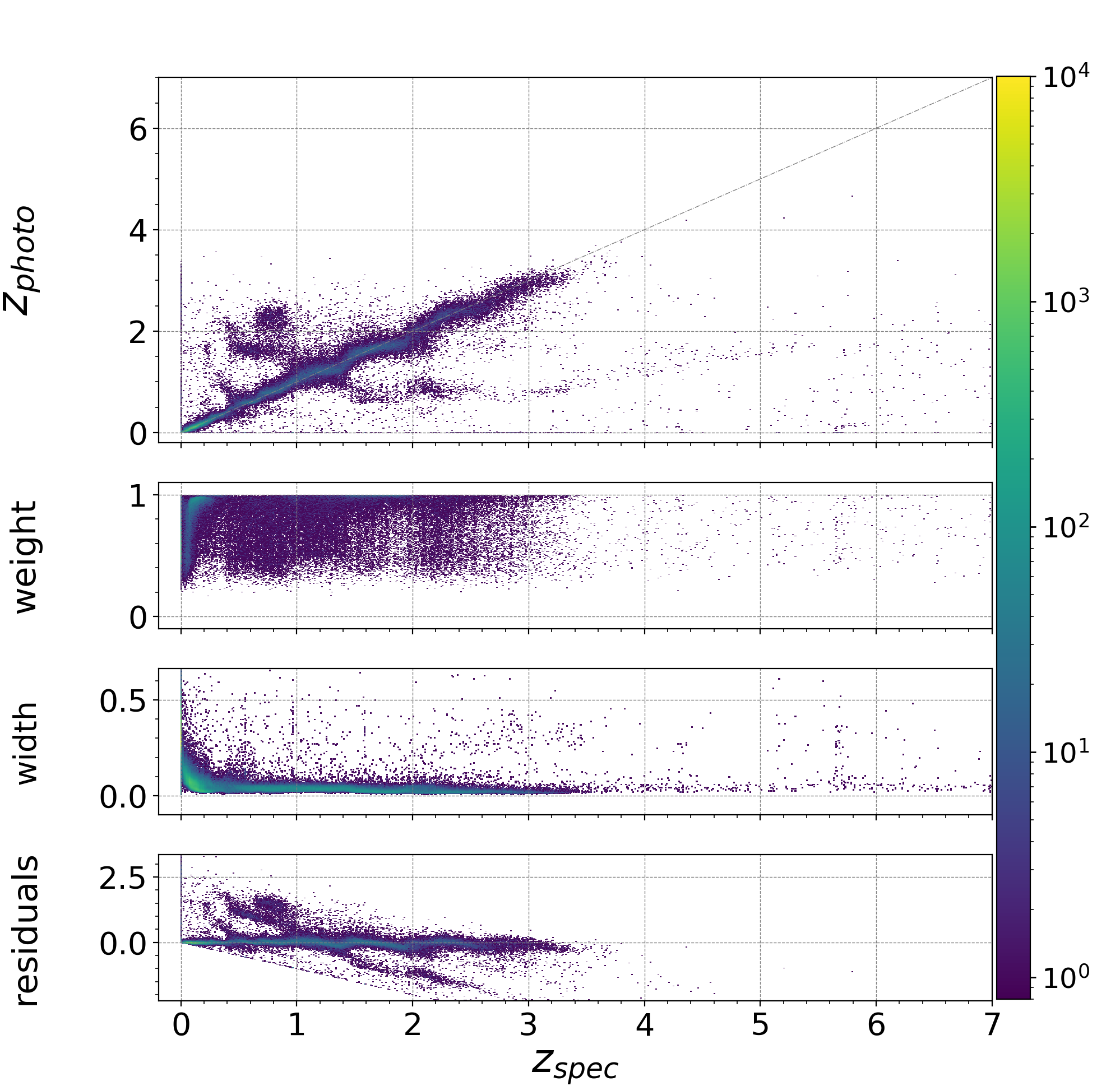}
    \includegraphics[width=0.43\textwidth]{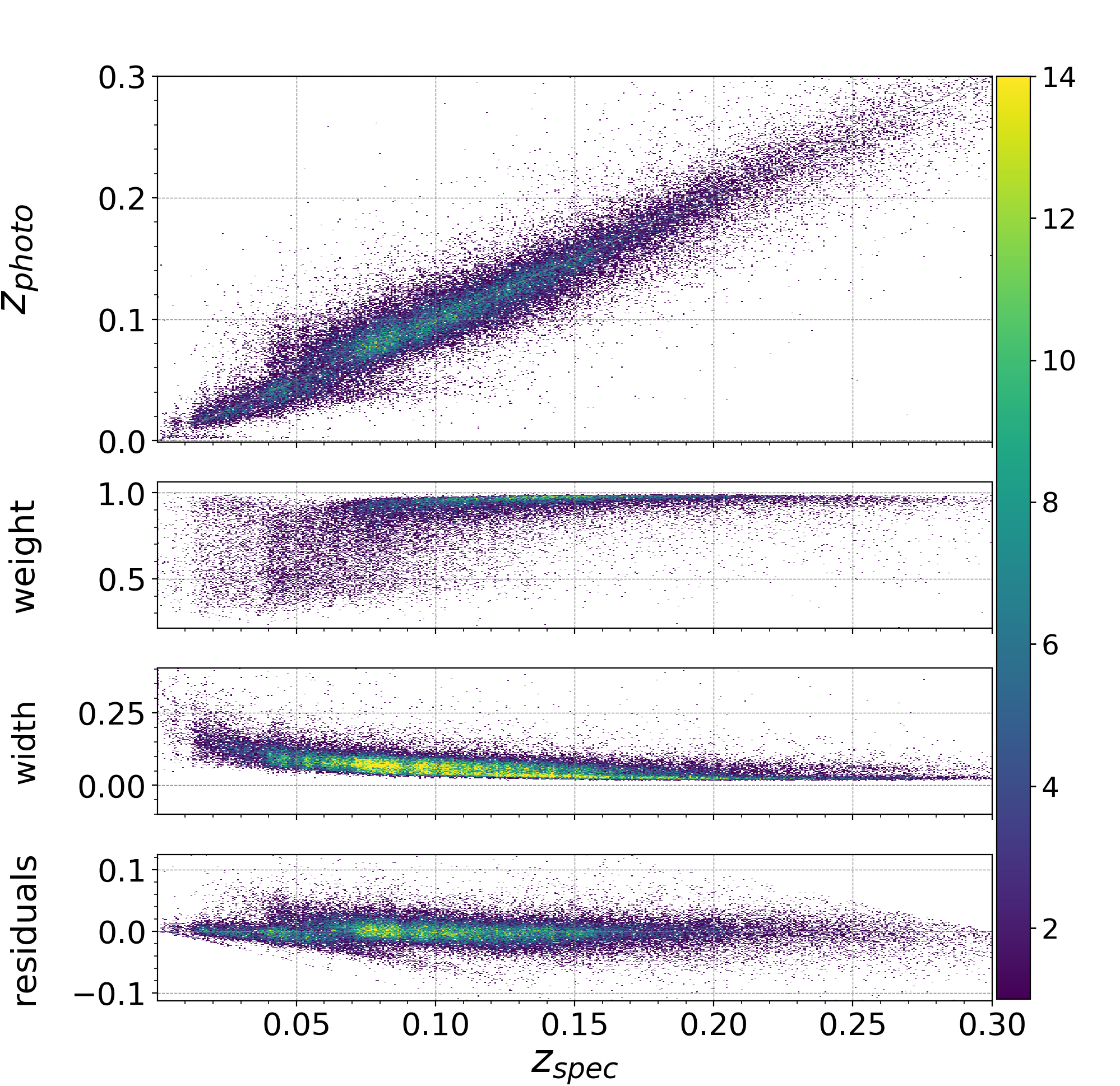}\\
    \includegraphics[width=0.43\textwidth]{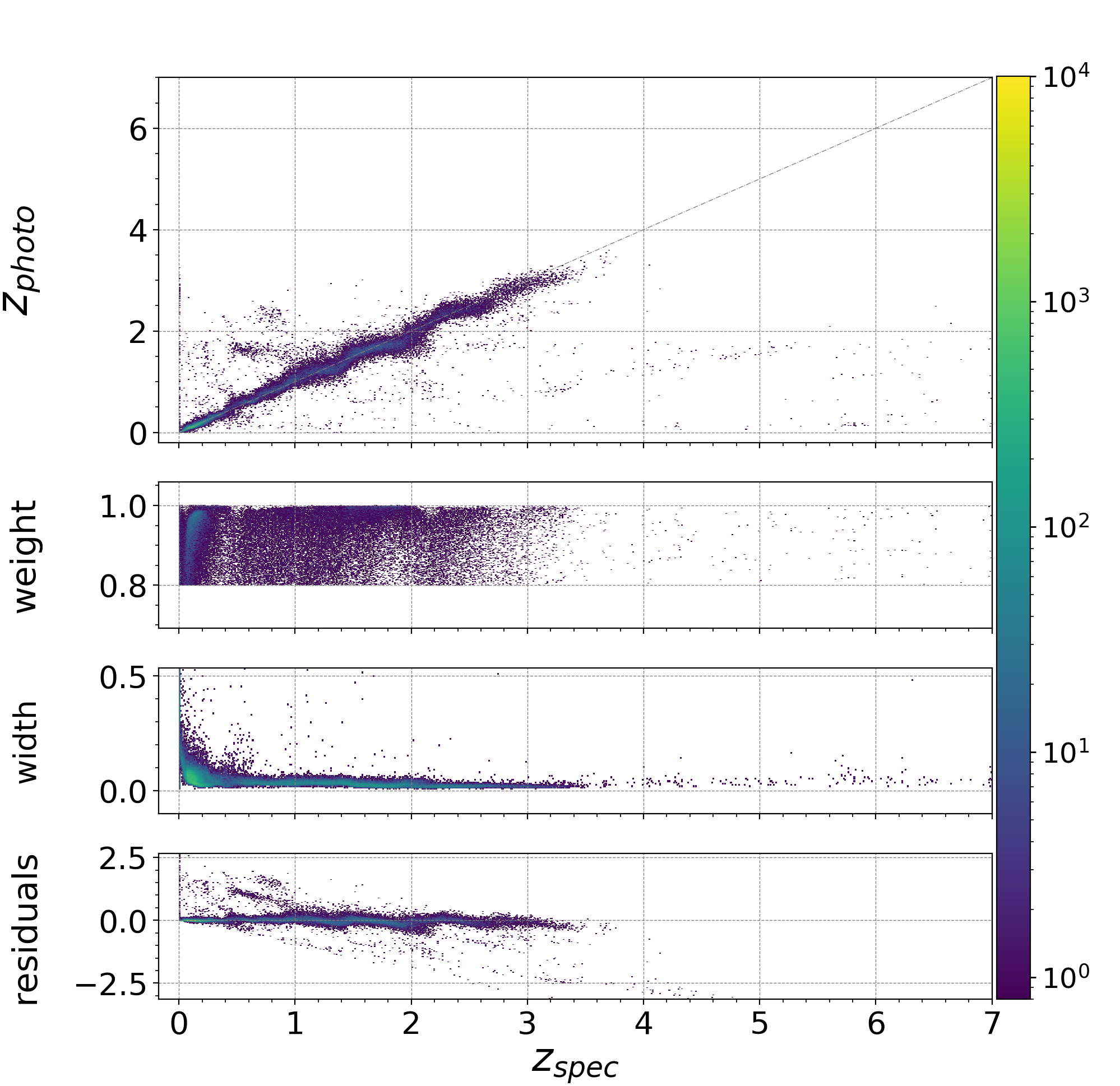}
    \includegraphics[width=0.43\textwidth]{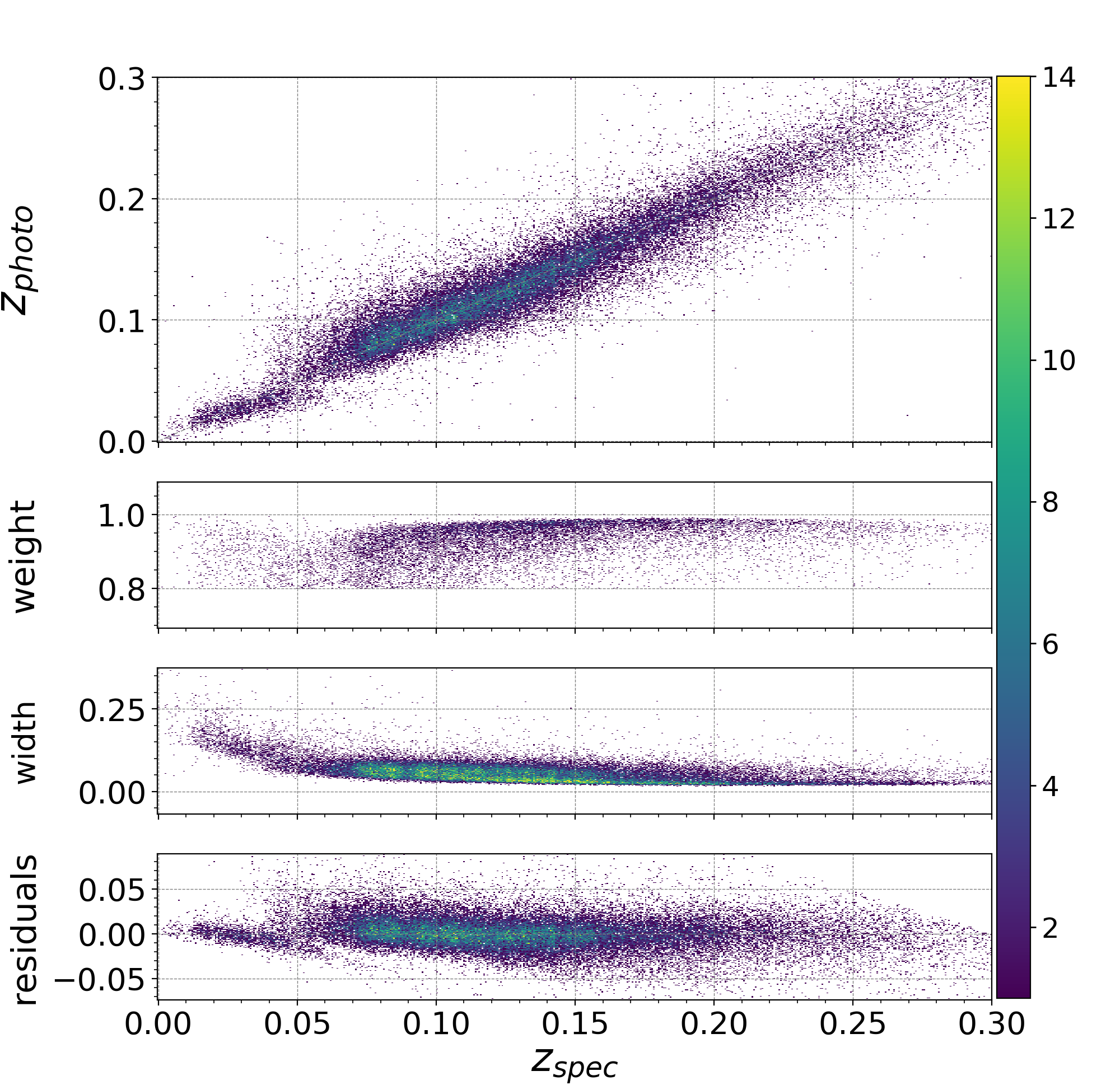}
    \caption{Photo-$z$ performance of different MDN implementations. \textit{Top panel:} Retaining only objects with $weight_{max} > 0.8$ membership probability to a MDN branch.
    \textit{Middle panel:} Including $u-$band \texttt{PSF} and \texttt{model} magnitudes. \textit{Bottom panle:} $u-$band magnitudes and MDN branch $weight_{max} > 0.8$.
    \textit{Right column:} All objects in the spectroscopic data-set. \textit{Left column:} Only spectroscopic galaxies in $z_{s}<0.3.$.
    }
    \label{fig:Photoz-specz-prob}
\end{figure*}

Figure~\ref{fig:Photoz-specz} shows the distribution of peak photo-$z$s (top) and expectation photo-$z$s (bottom) versus spectroscopic redshifts, $z_{s}$, for the MDN run with ten branches. 
One aspect to consider when determining photo-$z$ in cosmological wide-field imaging surveys, is the availability of $u-$band magnitudes, which is currently available for KiDS but not for DES. The \textit{Rubin} LSST is expected to deliver $u-$band photometry at the same depth of KiDS over $\approx 30\,000\mathrm{deg}^{2}.$ 
To test the effect, we re-trained one of our mixture models (\textit{IGMM spec. class}) for a data-set that includes $u-$band \texttt{PSF} and \texttt{model} magnitudes as additional input features (Fig.~\ref{fig:Photoz-specz-prob}). The bias and root-mean-square residuals are provided in Table~\ref{table:performance} for all objects and for galaxies with spectroscopic redshifts $z_{s}<0.3$, $z_{s}<0.4$, and $z_{s}<0.5$. This test leads to a lower rms $\Delta z$ and smaller fraction of $3\sigma$ outliers than for the same model without $u$-band magnitudes and can be considered as an improvement in accuracy.
Furthermore, with respect to the cross-contamination problem, this model also improves the overall confidence level with which an object belongs to a branch. As demonstrated in  Fig.~\ref{fig:Photoz-specz-prob}, bottom panel, the MDN performs ideed better for objects with increased confidence level.  

\begin{table}
\caption{Comparison between the photo-$z$ evaluation on all objects from the spectroscopic samples and the available SDSS photo-$z$s.
}
\label{table:comparisonphotoz_afterW}      
\centering          
\begin{tabular}{p{2cm}|p{1.5cm}|p{1.5cm}|p{2cm}} 
 & Bias & rms & $3\sigma$ outliers
  \\
  \hline\hline
  SDSS & -0.0038 & 0.0571 & 0.28\%
    \\
    \hline
    spec. class + $griz,$ $W1,$ $W2$ & -0.0003 & 0.0503 &  0.24\%
    \\
    \hline
\end{tabular}
\end{table}

\section{Discussion}
\label{Discussion}

\begin{figure}
\centering
\includegraphics[width=1\hsize]{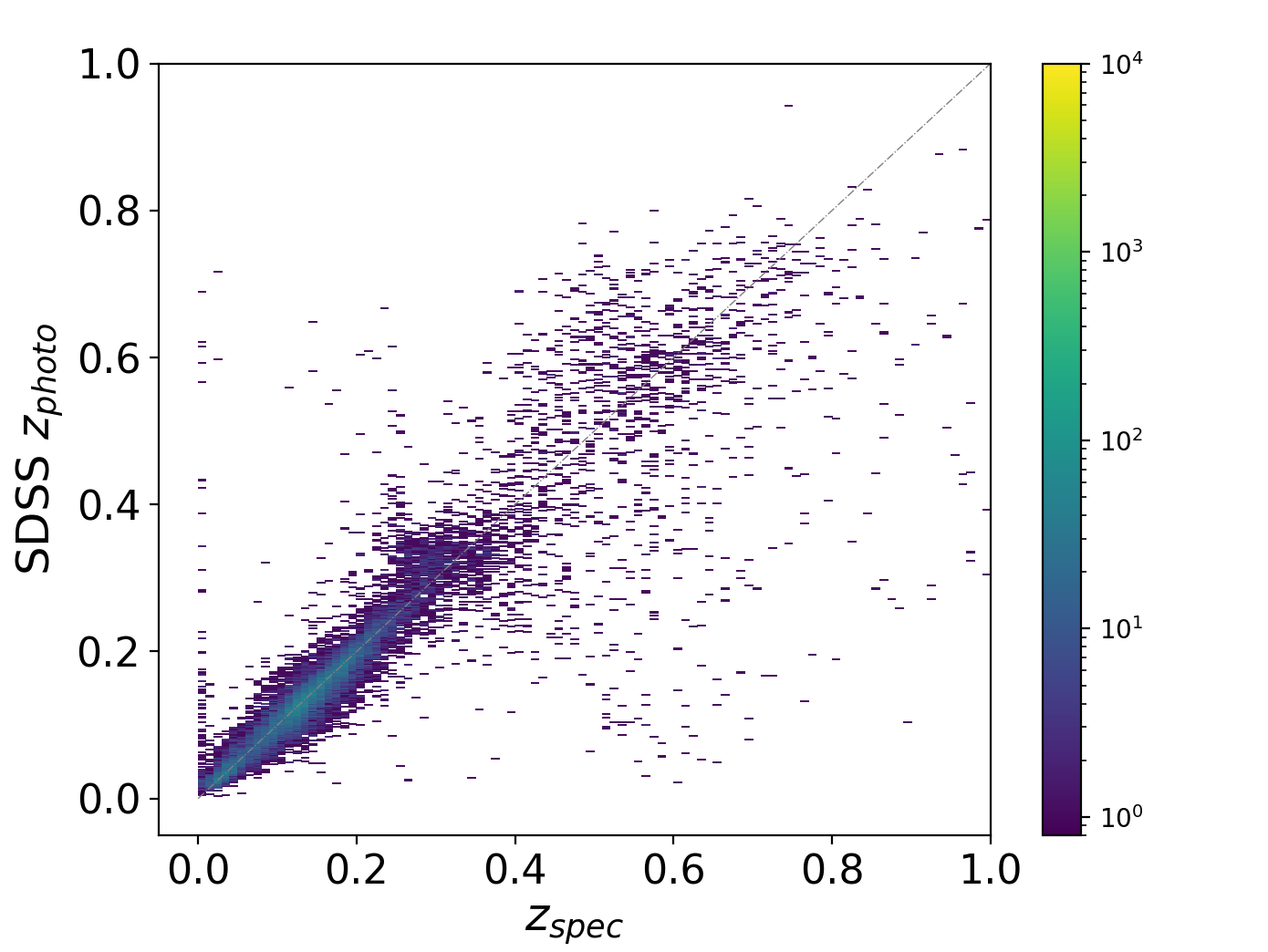}
\\
\includegraphics[width=1\hsize]{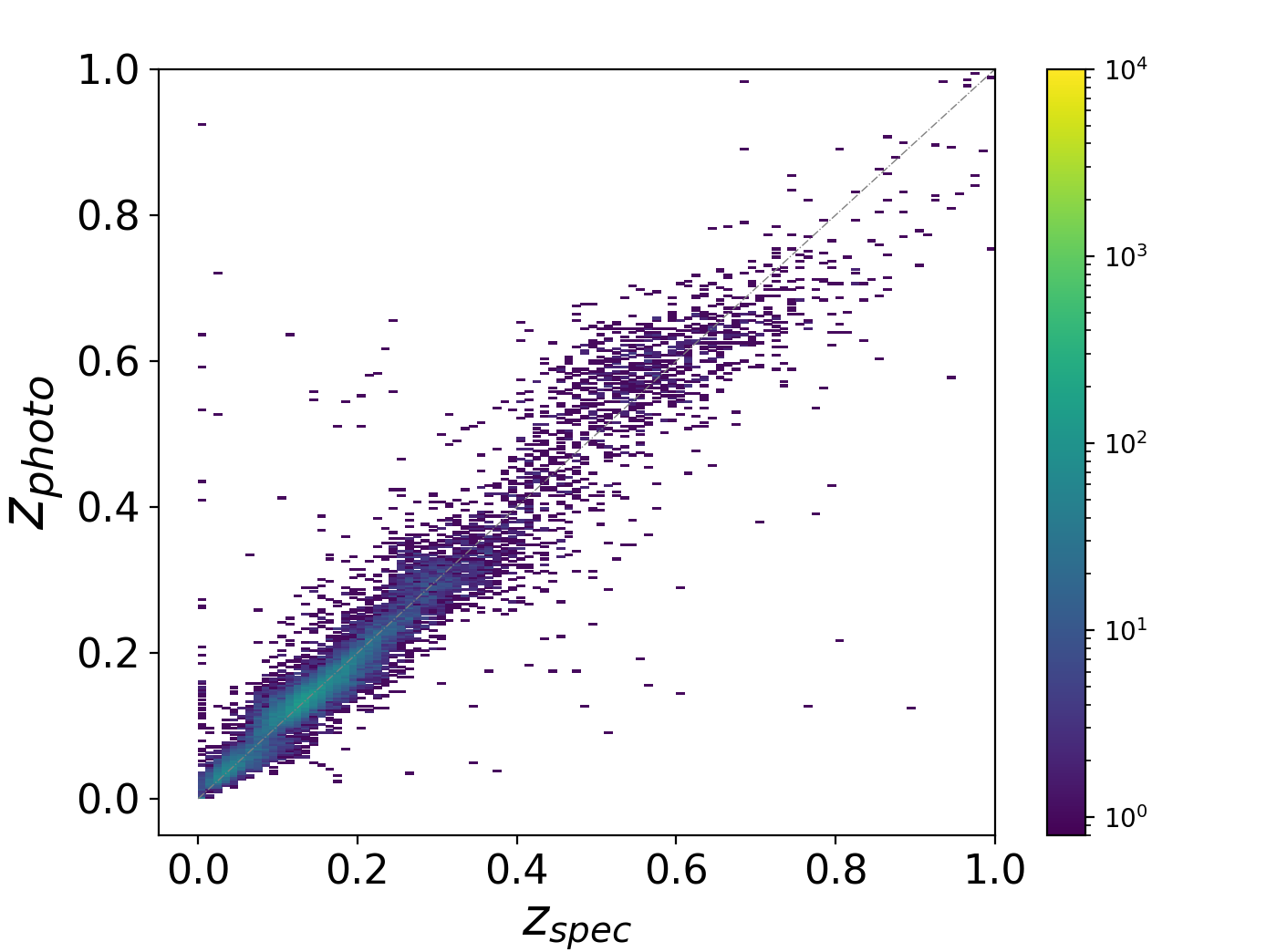}
\caption{\textit{Top panel:} SDSS spectroscopic redshift vs. SDSS photometric redshift. \textit{Bottom panel}: spectroscopic redshift vs. photometric redshift (this work). Colour bars indicate the number of sources in the diagrams.
The selection of sources is made by retaining objects with $weight_{max} > 0.8$ membership probability to a MDN branch.
}
\label{fig:comparisonphotoz}
\end{figure}

Table~\ref{table:comparisonphotoz_afterW} and Figure~\ref{fig:comparisonphotoz} show a comparison of our MDN peak photo-$z$ with those from the SDSS, which were obtained with a \textit{kNN} interpolation (18\,355 sources). All metrics are improved, with the added advantage that the MDN computes photo$-z$s for all objects (instead of just those with low stellarity) and can also cover the $z_{s}>1$ range more accurately than the SDSS \textit{kNN}. As a matter of fact, the SDSS photo-$z$s hardly exceed $z_{p}\approx1$, while our machinery is trained over a much wider redshift range.

\begin{table*}
\caption{MDN performance evaluation exclusively for sources with MDN branch $weight_{max} > 0.8$. The bias and rms are computed
 using the definition of clipped bias and rms in PS1-STR \citep{2020MNRAS.tmp.2033B}.
}
\label{table:performance-8-nonout}   
\centering 
\begin{tabular}{p{2.5cm}|p{2.4cm}|p{1.5cm}|p{1.5cm}|p{1.6cm}|p{1.6cm}|p{1.6cm}|p{1.6cm}} 
  IGMM 
  & photometry 
  & $\langle\Delta z\rangle$
  & rms($\Delta z$)  
  & $\langle\Delta z\rangle,$
  &  rms($\Delta z$),
  & rms($\Delta z$),
  &rms($\Delta z$),\\
  implementation
  &
  &($z_{s}<1$)
  &($z_{s}<1$)
  &range1\tablefootmark{a}
  &range1\tablefootmark{a}
  &range2\tablefootmark{b}
  &range3\tablefootmark{c}
  \\
  \hline\hline 
  Fully unsup. & $griz,$ $W1,$ $W2$ & 0.0005 & 0.0223  & 0.0003 & 0.0169  & 0.0192 & 0.0201\\
  \hline 
  spec.~class & $griz,$ $W1,$ $W2$  & 0.0007 & 0.0238 & $9\times 10^{-5}$ & 0.0153 & 0.0201 & 0.0209 \\
  \hline 
  redshift ($z_{s}$) & $griz,$ $W1,$ $W2$  & 0.0007 & 0.0217 & -0.0013 & 0.0167  & 0.0185 & 0.0196 \\
  \hline 
  spec.~class, $z_{s}$ & $griz,$ $W1,$ $W2$  & -0.0014 & 0.0235 & 0.0029 & 0.0167  & 0.0195 & 0.0197\\
  \hline 
  spec.~class & $ugriz,$ $W1,$ $W2$  & 0.0008 & 0.0186 & 0.0001 & 0.0148 & 0.0165 & 0.0169\\
  \hline
\end{tabular}
\tablefoot{
\tablefoottext{a}{Restricted to galaxies with $z_{s}<0.3$; } 
\tablefoottext{b}{Restricted to galaxies with $z_{s}<0.4$; } 
\tablefoottext{c}{Restricted to galaxies with $z_{s}<0.5$\ .}
}
\end{table*}
As a general benchmark, the LSST system science requirements document \footnote{https://docushare.lsstcorp.org/docushare/dsweb/Get/LPM-17} defines three photometric redshift requirements for a sample of four billion galaxies with $i<25$ mag within $z_{s} < 0.3$ as follows:
\begin{itemize}
    \item the rms$(\Delta z)< 0.02$ for the error in (1 + $z_{s}$) 
    \item the fraction of $3\sigma$ ("catastrophic") outliers < 10\%
    \item bias < 0.003
\end{itemize}
In our approach, these requirements are met if the MDN peak $z_{p}$ is adopted. The rms $\Delta z$ can be brought to 0.02 over $0<z_{s}<0.5$ if we restrict to "high-confidence" objects with $>0.8$ membership probability to a branch (Table~7; called \textit{weight} in Sect.~\ref{ss:Mixture Density Network}). Recently, \citet{Beck2020} used neural networks to classify objects in the Pan-STARRS1 footprint, which is known to have a more accurate photometry than the SDSS \citep{Magnier2013}, and evaluated photo-$z$s on objects with a probability $p>0.8$ of being galaxies, obtaining rms($\Delta z$)=0.03 over $0<z_{s}<1$. If we follow the same definitions and clipping\footnote{Their clipping procedure removes objects with $|\Delta z|>0.15$.} as by \citet{Beck2020}, then we obtain 1.7-2\% relative rms over the $0<z_{s}<0.5$ redshift range. Adding $u-$band information, as is the case with the SDSS and will be the case with the LSST, reduces the bias and fraction of outliers in all the redshift ranges considered. This is also because adding $u-$band magnitudes sharpens the MDN separation into branches and increases the fraction of objects with the highest weighted branch $>0.8$, as can be seen in the bottom panels of Figure~8.

We remark that throughout this work, we are simply adopting reddening values in the $i-$band ($A_{i}$), which the SDSS provides via a simple conversion of measured $E(B-V)$ values with a Milky-Way extinction law and $R_{V}=3.1.$ Our approach accounts for the systematic uncertainties due to the unknown extinction law by producing probability distributions and associate uncertainties for each photo-$z$ value.

The combined information across the optical and infrared, through the SDSS and WISE magnitudes, helps reducing the overlap between different classes in colour-magnitude space. The WISE depth is not a major limiting factor in the sample completeness as long as samples from the SDSS are considered, but it can affect the completeness significantly for deeper surveys \citep{SpiniAgn19}. In view of performing the classification and photo-$z$ estimation on the DES, and on the \textit{Rubin} LSST later on, deeper mid-IR data are needed. The \textit{unWISE} reprocessing of the WISE cutouts improved upon the original WISE depth \citep{Lang2014}. Further in the future, forced photometry of the unWISE cutouts from wide-field optical and NIR surveys may further increase the mid-IR survey depth \citep[e.g.][]{Lang14b}.

In general, separating objects into many sub-classes aids the photo-$z$ regression, as each MDN branch only needs to consider a subset of objects with more homogeneous properties than the whole photometric sample. Furthermore, the approach that we used in this work is both in the realm of machine learning (hence less constrained by choices of templates) while it can also produce a full output distribution for the photo$-z$ given the available photometric information. Beyond their first implementation in this work, mixture models can be easily adapted so that they can account for missing entries and limited depth, as in the GMM implementation by \citet{Melchior2018}.

\begin{acknowledgements}
    This work is supported by a VILLUM FONDEN Investigator grant (project number 16599) and Villum Young Investor Grant (project number 25501). This project is partially funded by the Danish council for independent research under the project ``Fundamentals of Dark Matter Structures'', DFF--6108-00570.
\end{acknowledgements}

\bibliographystyle{aa} 
\bibliography{main.bbl} 

\begin{appendix}
\label{app:a}
\section{IGMM}
Probability density distribution (PDF) formalization by Gaussian mixture modeling for K components is defined as follows:
\begin{equation}
    P(x|\mu_{1},...,\mu_{K},\Sigma_{1},...,\Sigma_{K})=\sum_{k=1}^{K} \pi_{k} \mathcal{N}(\mu_k, \sigma_{k})
\end{equation}
where $x$ is the data, $\pi_{k}$ is the weight distribution of mixtures that is defined by a Dirichlet distribution and $\sum_{k=1}^{K}=1 \pi_{k}$.

IGMM is the GMM case with infinite number of components using Dirichlet process instead of Dirichlet distribution to define the prior over the mixture distribution. Dirichlet process is a distribution over distributions, parameterizing by concentration parameter $\alpha$ and a base distribution $G_{0}$. The base distribution is the Dirichlet distribution which is a prior over the locations of components in the parameter space (i.e. $\Theta=(\mu, \Sigma)$). The concentration parameter $\alpha$ expresses the strength of belief in $G_{0}$ and affects the components weight \citep{Gorur:2010aa}.\\
Based on Bayes rule:
\begin{equation}
\label{eq:posterior cal}
    \gamma Z_{i}(k)=P(Z_{i}=k|x)= \frac{P(k)P(x|Z_{i}=k)}{P(x)}=\frac{\pi_{k} \mathcal{N}(x|\Theta_{k})}{\sum_{k=1}^{k} \pi_{k} \mathcal{N}(x|\Theta_{k})}
\end{equation}
where $\underline{\pi}$ is considered as the Dirichlet process and $Z_{i}$ is the latent variable. $\pi_{k}= N_{k}/N$ represents the effective number of data points assigned to the k-th mixture component. Despite the fact that we do not know the latent variable, there is information about it in the posterior.

Using an expectation-maximization (EM) algorithm to find the maximum likelihood with respect to the model parameters includes two steps, estimation step (e-step) and maximization step (m-step). After initializing the model parameters and evaluating the log-likelihood, the e-step evaluates the posterior distribution of $Z_{i}$ using the current model parameter values by equation~(\ref{eq:posterior cal}). Then, the m-step updates the model parameters based on the calculated latent variable as follows:
\begin{equation}
    \mu_{k} =\frac{\sum_{i=1}^{N} \gamma Z_{i}(k) x_{i}}{\sum_{i=1}^{N} \gamma Z_{i}(k)}=\frac{1}{N_{k}} \sum_{i=1}^{N} \gamma Z_{i}(k) x_{i}
\end{equation}
\begin{equation}
    \Sigma_{k} = \frac{1}{N_{k}} \sum_{i=1}^{N} \gamma Z_{i}(k) (x_{i} - \mu_{k})(x_{i} - \mu_{k})
\end{equation}
\begin{equation}
    \pi_{k} = \frac{N_{k}}{N} \text{ where } N_{k}=\sum_{i=1}^{N} \gamma Z_{i}(k)
\end{equation}

Eventually, the algorithm detects the convergence by the lack of significant change in the log-likelihood value from one iteration to the next, using:

\begin{equation}
    \log P(x|\mu, \Sigma, \pi)= \sum_{i=1}^{N} \log \Bigg\{ \sum_{k=1}^{K}
    \pi_{k} \mathcal{N}(x_{i}|\mu_{k}, \Sigma_{k}) \Bigg\}
\end{equation}
where $\pi_{k}$, the mixture proportion,  represents the probability of $x_{i}$ belonging to the k-th mixture component.

\end{appendix}

\end{document}